\documentclass[prx,
               twocolumn,
               english,
               superscriptaddress,
               floatfix,
               longbibliography, amsmath,amssymb]{revtex4-2}

\usepackage{graphicx}
\usepackage[caption=false]{subfig}
\usepackage{dcolumn}
\usepackage{bm}
\usepackage{braket}
\usepackage{appendix}
\usepackage{todonotes}
\usepackage{soul}
\usepackage[colorlinks = true, linkcolor = blue, citecolor = blue, urlcolor = blue]{hyperref}
\usepackage{academicons}
\usepackage{multirow}

\usepackage{lipsum}

\newcommand{\orcid}[1]{\href{https://orcid.org/#1}{\textcolor[HTML]{A6CE39}{\aiOrcid}}}
\newcommand{\aSi}{\mbox{a-Si}}

\newcommand{\andrea}[1]{%
  \bgroup
  \hskip0pt\color{red!80!black}%
  #1%
  \egroup
}

\newcommand{\pete}[1]{%
  \bgroup
  \hskip0pt\color{green!80!black}%
  #1%
  \egroup
}

\newcommand{\B}{$B_{3\text{dB}}$}

\begin{document}

\preprint{APS/123-QED}

\title{Kinetic Inductance Traveling Wave Parametric Amplifiers Near the Quantum Limit: Methodology and Characterization}

\author{L.~Howe}
\email{logan.howe@nist.gov}
\email{howel@colorado.edu}
\affiliation{National Institute of Standards and Technology, Boulder, 80305, Colorado, USA}
\affiliation{Department of Physics, University of Colorado, Boulder, 80309, Colorado, USA}
\affiliation{Amazon Web Services Center for Quantum Computing, Pasadena, 91106, California, USA}

\author{A.~Giachero}
\email{andrea.giachero@colorado.edu}
\affiliation{National Institute of Standards and Technology, Boulder, 80305, Colorado, USA}
\affiliation{Department of Physics, University of Colorado, Boulder, 80309, Colorado, USA}
\affiliation{Department of Physics, University of Milano Bicocca, Milan, I-20126, Italy}

\author{M.~Vissers} 
\affiliation{National Institute of Standards and Technology, Boulder, 80305, Colorado, USA}

\author{P.~Campana}
\affiliation{Department of Physics, University of Milano Bicocca, Milan, I-20126, Italy}
\affiliation{INFN - Milano Bicocca, Milan, I-20126, Italy}

\author{J.~Wheeler} 
\affiliation{National Institute of Standards and Technology, Boulder, 80305, Colorado, USA}


\author{J.~Gao} 
\affiliation{National Institute of Standards and Technology, Boulder, 80305, Colorado, USA}
\affiliation{Department of Physics, University of Colorado, Boulder, 80309, Colorado, USA}
\affiliation{Amazon Web Services Center for Quantum Computing, Pasadena, 91106, California, USA}

\author{J.~Austermann} 
\affiliation{National Institute of Standards and Technology, Boulder, 80305, Colorado, USA}

\author{J.~Hubmayr}
\affiliation{National Institute of Standards and Technology, Boulder, 80305, Colorado, USA}

\author{A.~Nucciotti}
\affiliation{Department of Physics, University of Milano Bicocca, Milan, I-20126, Italy}
\affiliation{INFN - Milano Bicocca, Milan, I-20126, Italy}

\author{J.~Ullom}
\affiliation{National Institute of Standards and Technology, Boulder, 80305, Colorado, USA}
\affiliation{Department of Physics, University of Colorado, Boulder, 80309, Colorado, USA}

\date{\today}

\begin{abstract}
We present a detailed simulation and design framework for realizing traveling wave parametric amplifiers (TWPAs) using the nonlinear kinetic inductance of disordered superconductors -- in our case niobium-titanium-nitride (NbTiN). These kinetic inductance TWPAs (KITs) operate via three-wave mixing (3WM) to achieve high broadband gain and near-quantum-limited (nQL) noise. Representative fabricated devices -- realized using an inverted microstrip (IMS), dispersion-engineered, artificial transmission line -- demonstrate power gains above 25~dB, bandwidths beyond 3~GHz, and achieve ultimate system noise levels of 1.1~quanta even when operated with no magnetic shielding. These performance metrics are competitive with state-of-the-art Josephson-junction-based TWPAs but involve simpler fabrication and able to providing three orders of magnitude higher dynamic range ($IIP_1 = -68$~dBm, $IIP_3 = -55$~dBm), and high magnetic field resilience -- making KITs an attractive technology for highly multiplexed readout of quantum information and superconducting detector systems.
\end{abstract}

\maketitle



\section{Introduction\label{sec:intro}}
The rapid growth of emergent quantum technologies in both research and industry -- as well as the large number of planned and fielded astronomy and fundamental physics instruments with large arrays of superconducting detectors -- increasingly calls for higher performing and more capable amplifiers. Ideally, these amplifiers operate near the quantum limit of added noise (QL). Resonant Josephson parametric amplifiers (JPAs) exploiting the nonlinear Josephson inductance have enjoyed wide success within the quantum computing \cite{kaufman2023josephson, mutus2014strong, nilsson2024small, wang2025high, walter2017rapid, abdo2021high, lecocq2021efficient} and dark matter communities \cite{backes2021quantum, quiskamp2025near, bai2025dark}. Indeed, JPAs continue to play a fundamental role in the development of quantum technologies however, transcending their narrow bandwidth ($\lesssim 100$~MHz) and inherent gain-bandwidth product limit \cite{metelmann2022quantum} to facilitate highly-frequency-division-multiplexed systems is difficult. Adopting a traveling wave parametric amplifier (TWPA) architecture widens the parametric amplifier bandwidth to a few gigahertz and makes high levels of multiplexing more feasible. Such Josephson TWPAs (JTWPAs) are particularly crucial for applications such as multiplexed qubit readout \cite{Heinsoo2018, remm2023intermodulation}, micro-oscillator measurements~\cite{Youssefi2023}, quantum key distribution \cite{Fesquet2022} and, more recently, in a variety of fundamental physics experiments \cite{Braggio2022, bartram2023dark, Navarro2021, Ramanathan2023wideband}. However, Josephson-junction-based parametric amplifiers are hampered by their complex design, involved fabrication, high sensitivity to magnetic fields, and very low power handling. Parametric amplifiers based on superconducting films with a high kinetic inductance offer promising solutions to these limitations and are thus an advantageous alternative Josephson-based amplifiers. Kinetic-inductance-based amplifiers require only a single nonlinear film (as opposed to thousands of Josephson junctions), routinely display dynamic ranges more than 30~dB higher \cite{malnou2021three, vissers2016low, faramarzi2024kinetic}, and, in the case of resonant amplifiers, have demonstrated resilience Tesla-scale magnetic fields \cite{vaartjes2024strong, frasca2024three, xu2023magnetic}.

To date one of most significant roadblocks in widespread adoption of kinetic inductance TWPAs (KITs) is the high pump powers historically needed in these devices -- sometimes as much as 10 orders of magnitude above levels required for superconducting qubit or resonator measurements for NbTiN-based devices \cite{ho2012wideband, vissers2016low}. These powers cause heating of components, prevent adequate thermalization of the pump input lines, and spoil the achievement of quantum-limited performance. In recent years significant progress has been made in reducing the KIT pump power by decreasing the cross-sectional area of the high kinetic inductance film. This also has the added benefit of shortening the length of transmission required to achieve sufficient gain and subsequently improves device yield. Noteworthy performance increases have been made by moving to a stub-loaded coplanar waveguide (CPW) geometry with a minimum feature size of 1~$\mu$m (typical optical lithography limit). In this geometry, as the transmission line characteristic impedance is $Z_0 = \sqrt{\mathcal{L}/\mathcal{C}}$, long quarter-wave stubs are added to increase the capacitance per unit length $\mathcal{C}$ to compensate for the large kinetic inductance per unit length $\mathcal{L}$ and create a matched (50~$\Omega$) transmission line impedance \cite{malnou2021three}. Periodic modulation of the stub lengths is then used to create a photonic bandgap and allow broadband phase-matching and gain over a desired frequency range. These dispersion engineering techniques have permitted the successful demonstration of KITs with bandwidths over 2~GHz and, when used as the first-stage amplifier (FSA) in a multi-stage readout chain, system noise levels of 3.1~quanta \cite{malnou2021three} -- competitive with all contemporary JTWPAs \cite{macklin2015near, qiu2023broadband, wang2025high, gaydamachenko2025rf, kaufman2023josephson, simbierowicz2021characterizing, malnou2024traveling, ranadive2024traveling, ranadive2022kerr}.

However, high-yield fabrication of stub-loaded CPW KITs with a total lengths above 30~cm \cite{malnou2021three} (required to meet the high $\sim 20$~dB gain target) is challenging due to persistent short-to-ground failures of the CPW center conductor. Recent adoption of an inverted microstrip (IMS) topology with a low-loss dielectric \cite{shu2021nonlinearity} bestows significant advantages. First, if the dielectric is sufficiently thick to be free of pinholes (in our case 100~nm), this eliminates the short-to-ground lithographical failure mode. Second, choosing a high-permittivity dielectric, such as amorphous silicon (\aSi), and working with thicknesses of a few hundred nanometers, increases $\mathcal{C}$ by nearly an order of magnitude. This shortens the stub length required to match to 50~$\Omega$ by the same factor and subsequently reduces the likelihood of shorts along the stubs while widening the gain bandwidth. Third, an elementary cell in a stub-loaded CPW has a minimum length of five squares while for the IMS it is two squares -- which reduces the required length of transmission line for high gain. Finally, the single, cohesive sky ground plane of the IMS topology presents a much cleaner rf grounding environment and reduces reflections and resonances across the device.

We have adopted the IMS topology for our NbTiN-based three-wave mixing KITs and have recently demonstrated an order of magnitude pump power reduction by reducing the NbTiN film thickness while maintaining a 1~$\mu$m minimum feature size \cite{giachero2024kinetic, howe2025compact}. In this article we detail the design and simulation of our IMS KITs which demonstrate excellent model-to-hardware correlation when incorporating basic properties of our fabrication platform (film kinetic inductance and specific capacitance). We present a detailed suite of diagnostic and tune-up measurements including: time domain reflectometry (TDR) extraction of the artificial transmission line (TL) characteristic impedance, input compression power (linear and cubic response), critical and scaling current extraction, and gain measurements at various pump frequencies and powers. Finally, we measure the system noise with a KIT as the readout chain FSA and demonstrate a meaningful reduction in noise and improved bandwidth relative to prior generation IMS and CPW KITs \cite{giachero2024kinetic, malnou2021three}.

\section{\label{sec:kit_physics}KIT Principle of Operation}
The kinetic inductance per unit length of a superconducting film as a function of current, $I$, is nontrivial \cite{kubo2020superfluid} but can be Taylor-expanded in the small-signal limit in even powers of $I$ \cite{anlage2002current}:
\begin{equation}\label{eq:kin0}
    \mathcal{L}_k(I) = \mathcal{L}_0 \left[ 1 + \left(\frac{I}{I_{*,2}}\right)^2 + \left(\frac{I}{I_{*,4}}\right)^4+\mathcal{O} \left(I^6\right) \right].
\end{equation}
$\mathcal{L}_0$ is the zero-bias and zero-frequency kinetic inductance per unit length, and $I_{*,2}$ ($I_{*,4}$) is the second (fourth) order \textit{scaling current}. In principle $I_*$ can be frequency-dependent but is generally not observed to be so. As we can see, the inductance nonlinearity only becomes significant when the currents in the film approach the scaling current. We may redefine $I$ as the sum of the dc and pump (rf) currents $I = I_{dc} + I_p$ and re-write Eq.~(\ref{eq:kin0}) as 
\begin{equation}\label{eq:kin1}
    \mathcal{L}_k(I) = \mathcal{L}_0\left[ 1 + I_p(\varepsilon + \xi I_p) + \mathcal{O}(I^3)\right].
\end{equation}
The prefactors $\varepsilon$ and $\xi$ respectively describe the relative strength of the three-wave-mixing (3WM) and four-wave-mixing processes (4WM):
\begin{equation}\label{eq:eps}
    \varepsilon = \frac{2 I_{dc}}{I_*^2 + I_{dc}^2}~, ~~~ \xi = \frac{1}{I_*^2 + I_{dc}^2}
\end{equation}
and demonstrate that a predominantly 3WM or 4WM film may be realized with the application of a dc bias current.

If we consider the telegrapher's equations for a long transmission line composed of a high kinetic inductance material obeying Eq.~(\ref{eq:kin1}) we may use the coupled mode equations (CMEs) to enforce harmonic balance (energy conservation and mixing) between the pump, signal, idler triplet frequencies $\omega_j \in \{p, s, i\}$ \cite{malnou2021three}. From the CMEs we can extract the 3WM phase matching condition 
\begin{equation}
    \Delta_\beta = \Delta_k + \frac{1}{8} \xi I_{p0}^2 (k_p - 2k_s - 2k_i),
\end{equation}
where $\Delta_k = k_p - k_s - k_i$, $k_j$ is the wavenumber for frequency $\omega_j$ exiting the device, and $I_{p0}$ is the pump current amplitude at the device input. If phase matching is exactly satisfied ($\Delta_\beta = 0$) the CMEs yield the current amplitude at position $x$ along the transmission line \cite{boyd2008nonlinear}:
\begin{equation}
    I_{s,i} = \cosh(g_3 x)I_{s0,i0};
\end{equation}
demonstrating the power gain, $G$, is exponential in device length when phase matching occurs
\begin{equation}\label{eq:gain_analytic}
    G(x) = \left| \frac{I_s(x)}{I_{s0}} \right|^2 = \cosh^2(g_3 x).
\end{equation}
Here $g_3 = \varepsilon I_{p0} \sqrt{k_s k_i} / 4$ is the 3WM element.

\subsection{Motivations for Higher Kinetic Inductance}
Eq.~(\ref{eq:gain_analytic}) can be used to estimate the maximum gain of a  device given realistic values. The pump-induced inductance modulation is, to first order in $I_p$,
\begin{equation}
    \delta_L = \varepsilon I_{p0} =  \frac{2 I_p I_{dc}}{I_{*,2}^2 + I_{dc}^2}.
\end{equation}
Further, take the degenerate parametric amplification point $k_s = k_i = k_p = \omega_p \sqrt{\mathcal{L}_d \mathcal{C}}$ with 
\begin{equation}\label{eq:Ld}
    \mathcal{L}_d = \mathcal{L}_0(1 + I_{dc}^2 / I_*^2)
\end{equation}
such that:
\begin{equation}\label{eq:gain_explicit}
    G(x) = \cosh^2\left( \frac{1}{4} \frac{I_{dc} I_p}{I_{*,2}^2 + I_{dc}^2} \omega_p \sqrt{\mathcal{L}_d \mathcal{C}} x \right).
\end{equation}
Here all we have done is explicitly write the 3WM mixing element in terms of parameters we may control in fabrication or experiment. However, this elucidates certain avenues for optimization: because $g_3$ describes the gain per unit length, maximizing this quantity enables reaching a gain target with the shortest possible device length. Consider devices with a gain target of 20~dB and $Z_0 = \sqrt{\mathcal{L / C}} = 50~\Omega$. For a fixed pump frequency, i.e. fixed target gain bandwidth, $g_3$ actually increases linearly with respect to $\mathcal{L}$ as the same magnitude capacitance must be added to maintain the 50~$\Omega$ match. Therefore a $50~\Omega$ amplifier with double the unit length inductance reaches 20~dB with half the TL length.

The remaining component of $g_3$ is $\delta_L$ which is primarily an experimental parameter rather than a design parameter. Clearly, maximizing $\delta_L$ occurs when $I = I_{dc} + I_p = I_*$. Generally, $I$ cannot exceed $I_*$ because this would cause a runaway breakdown of superconductivity in the NbTiN film \cite{zmuidzinas2012superconducting}. In this case $\delta_L$ ranges from near zero (for $I_{dc} \sim 0$ and $I_c$) to 0.21 for $I_{dc} = 0.4 I_*$. In practice our device critical currents, $I_c$, (the dc current at which superconductivity is broken) are limited to approximately $I_* / 3$. This yields a maximum for $\delta_L$ of 0.027 when $I_{dc} \sim I_c = 0.16 I_*$. From this perspective, realizing devices whose critical currents are a much larger fraction of the scaling current would enable further reduction in device length by more than a factor of ten while still meeting the 20~dB gain target.

\begin{figure}[t]
    \centering\includegraphics[width=0.98\linewidth]{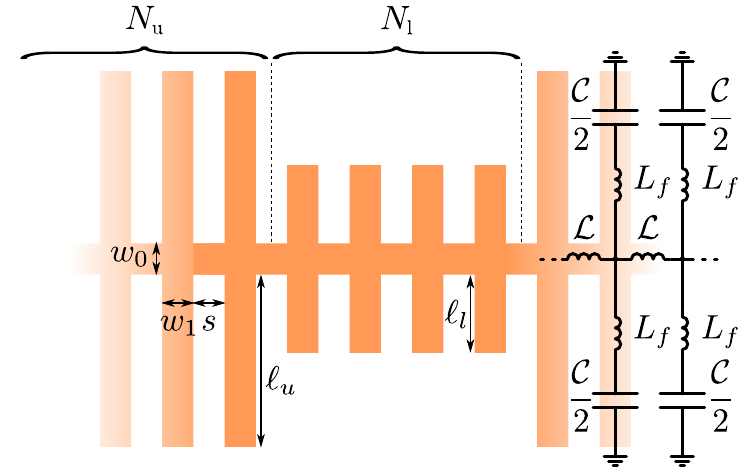}
    \caption{Schematic of the KIT amplifier stub-loaded artificial TL (not to scale). The highly inductive central line is loaded with stubs of length $\ell_u$ and $\ell_l$ which are tuned to achieve the desired impedance. $N_u=30$ longer stubs, of length $\ell_u$, form the \textit{unloaded} part of the supercell with a $48~\Omega$ impedance, while $N_l=4$ shorter stubs, of length $\ell_l$, form the \textit{loaded} part of the supercell with a $78~\Omega$ impedance. This combination creates a photonic stopband dispersion feature at the desired frequency and is given by the electrical length of the series of unloaded cells. In the equivalent electrical circuit, each cell consists of a series inductance $\mathcal{L}$ and both a parallel inductance $L_f$ and parallel capacitance $C/2$, for stubs on each side of the center line, to ground. $L_f$ and $C/2$ form a quarter-wave resonator at high frequency whose impact on the dispersion relation is negligible in our architecture. Typical values of $L_f \sim 0.4$~nH and $C \sim 10$~fF yield a stub resonance of 250~GHz, indicating for $\omega / 2 \lesssim 25$~GHz these discrete features are well within the lumped element limit.} 
    \label{fig:ims}  
\end{figure}

\section{Design and Simulation}\label{sec:design_and_simulation}
Our design process follows the formalism constructed in \cite{malnou2021three} relevant to stub-loaded artificial TLs with periodic $Z_0$ modulation dispersion engineering. To fully simulate our devices and obtain over 20~dB of gain at the desired frequency, and over the maximum bandwidth, we construct the full device cascaded $ABCD$ matrix and solve the CMEs. We apply realistic dc bias and pump amplitudes and sweep the pump frequency to verify the gain properties. However, the $ABCD$ matrix for each elementary cell requires inputs that must first be extracted from electromagnetic simulations of a small segment of the stub-loaded IMS TL. Notably, for each elementary cell, these are the series inductance $\mathcal{L}$, the inductance of the stub \textit{finger} loadings $L_f$, and the total unit cell capacitance to ground $\mathcal{C}$. Fig.~\ref{fig:ims} shows a schematic representation of the metal layer geometry and corresponding electrical circuit.

We perform single-port simulations of 320 unit cells with $w_0 = s = w_1 = 1~\mu$m constructed from a 10~nm thick NbTiN film with an expected $\mathcal{L}$ of 30~pH/$\square$ \cite{Giachero2023characterization}. By interchangeably leaving open and shorting the end of the TL to ground, we may predict $\mathcal{C}$ and $\mathcal{L}$, respectively, as a function of the length of the stubs. Besides yielding necessary parameters for the elementary cell $ABCD$ matrix, these simulations also inform the stub lengths required to create a unit cell with a specific characteristic impedance.

In the low frequency limit, and for the shorted-end boundary condition, the input inductance, $L_1$, of the IMS TL composed of $n = 320$ cells is
\begin{equation}\label{eq:fitL}
L_1\simeq
\mathcal{L}  n\left[1+
\cfrac{\omega^2  n^2}{3}  \mathcal{L}  \mathcal{C}+
    \cfrac{\omega^4  n^4}{15}  \mathcal{L}^2  \mathcal{C}^2 
\right],
\end{equation}
where $\omega$ is the frequency. For the open-end boundary condition the input capacitance, $C_1$ is
\begin{equation}\label{eq:fitC}
C_1\simeq
\mathcal{C}  n\left[
\cfrac{45}{45-15  \omega^2  n^2  \mathcal{L}  \mathcal{C}-\omega^4  n^4  \mathcal{L}^2  \mathcal{C}^2}
\right].
\end{equation}
See Appendix~\ref{sec:app:model} \footnote{Commercial instruments and software are identified in this paper in order to adequately specify the experimental procedure. Such identification does not imply recommendation or endorsement by NIST, nor does it imply that the product identified is necessarily the best available for the purpose.} for a detailed derivation of Eq.~(\ref{eq:fitL}) and Eq.~(\ref{eq:fitC}).

\begin{figure}[t]
    \centering
    \includegraphics[width=0.48\textwidth]{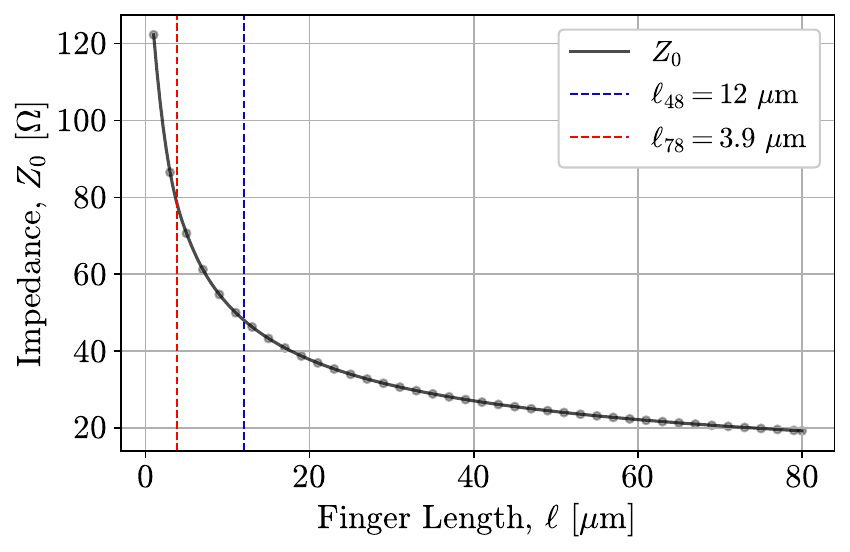}
    \caption{Unit cell impedance simulation results as a function of finger length. $\mathcal{L}$ and $\mathcal{C}$ are extracted using the shorted and open ended boundary conditions described by Eqs.~(\ref{eq:fitL}) and (\ref{eq:fitC}) and the 320 unit cell subsegment of TL. We use a 100~nm a-Si layer with a typical dielectric constant of $_r = 9.1$ and a NbTiN film with $L_0 = 30~$pH/$\square$.}
    \label{fig:Z0_vs_finger_length}
\end{figure}

Thus simulations at low frequency ($\omega / 2 \pi \leq 100$~MHz) of the input admittance, and their subsequent fits according to Eqs.~(\ref{eq:fitL}) and (\ref{eq:fitC}), yield $\mathcal{L}$ and $\mathcal{C}$ as a function of the stub length, $\ell$. While only one case is needed, we perform both sets of simulations and fits to ensure they are self-consistent. Our large feature size (1~$\mu$m) relative to the dielectric thickness (100~nm) places our devices strongly in the parallel plate capacitor limit where the unit cell capacitance is given by $C = \epsilon_0 \epsilon_r a / d$. $a$ is the unit cell total area (determined by $\ell,~w_0,~w_1$ and $s$ as in Fig.~\ref{fig:ims}) and $d$ is the dielectric thickness. The unit cell series inductance is simply $L = n_\square L_0$, with $n_\square$ the number of squares in the unit cell and $L_0$ the zero-bias and zero-frequency sheet kinetic inductance. Our simulations corroborate these assumptions and more details can be found in Appendix~\ref{sec:app:model}. For $L_0 = 30$~pH/$\square$ and $\epsilon_r = 9.1$ we obtain the $Z_0$ as a function of $\ell$ curve shown in Fig.~\ref{fig:Z0_vs_finger_length} -- with the stub lengths of 12.1~$\mu$m (3.9~$\mu$m) yielding $Z_0$ of $48~\Omega$ ($78~\Omega$).

We choose unit cell impedances of $48~\Omega$ and $78~\Omega$ so that under dc bias the increase in inductance forces the unit cell impedances to approach $50~\Omega$ and $80~\Omega$ -- although in principle these impedances can increase by up to 30\% depending on how close the device is biased to $I_c$. The artificial TL $Z_0$ is expected to be a weighted average of the unloaded and loaded cell impedance, however, in practice when $N_u \gg N_l$ both the weighting and the fact that the loaded cell sections are sufficiently sub-wavelength drives the effective TL $Z_0$ to $Z_u$.

With the geometry and expected electrical properties of the unit cells necessary for dispersion engineering known, we may proceed to simulate the response of a full device under arbitrary dc bias. The $ABCD$ matrix for a single unit cell at frequency $\omega$ is \cite{malnou2021three}
\begin{equation}\label{eq:abcd_general}
    \mathbf{T}_{u,l} = \begin{bmatrix} 1 & i \mathcal{L}_d \omega \\[1mm] \dfrac{i 2 \mathcal{C}^{u,l} \omega}{2 - L_f^{u,l} \mathcal{C}^{u,l} \omega^2} & 1 - \dfrac{2 \mathcal{L}_d \mathcal{C}^{u,l} \omega^2}{2 - L_f^{u,l} \mathcal{C}^{u,l} \omega^2} \end{bmatrix}.
\end{equation}
As shown in Fig.~\ref{fig:ims}, dispersion engineering takes the form of modulating the unit cell $Z_0$ via tuning the length of stubs. This modifies both $\mathcal{C}$ and $L_f$ so the transfer matrix $T_u~(T_l)$ for an unloaded (loaded) cell is given by Eq.~(\ref{eq:abcd_general}).

Combining $N_u$ unloaded and $N_l$ loaded unit cells symmetrically about the loaded cells gives the supercell transfer matrix
\begin{equation}
    \mathbf{T}_\text{sc} = \mathbf{T}_u^{N_u / 2} \mathbf{T}_l^{N_l} \mathbf{T}_u^{N_u / 2},
\end{equation}
and the whole device transfer matrix is
\begin{equation}\label{eq:full_dev_abcd}
    \mathbf{T}_\text{KIT} = \mathbf{T}_\text{sc}^{N_\text{sc}},
\end{equation}
where $N_\text{sc}$ is the number of supercells. The necessary product of this construction is the full KIT scattering parameters, which are related to and extracted from the full device $ABCD$ transfer matrix of Eq.~(\ref{eq:full_dev_abcd}) \cite{pozar2012microwave}. 

\begin{figure*}[t]
    \centering
    \includegraphics[width=\textwidth]{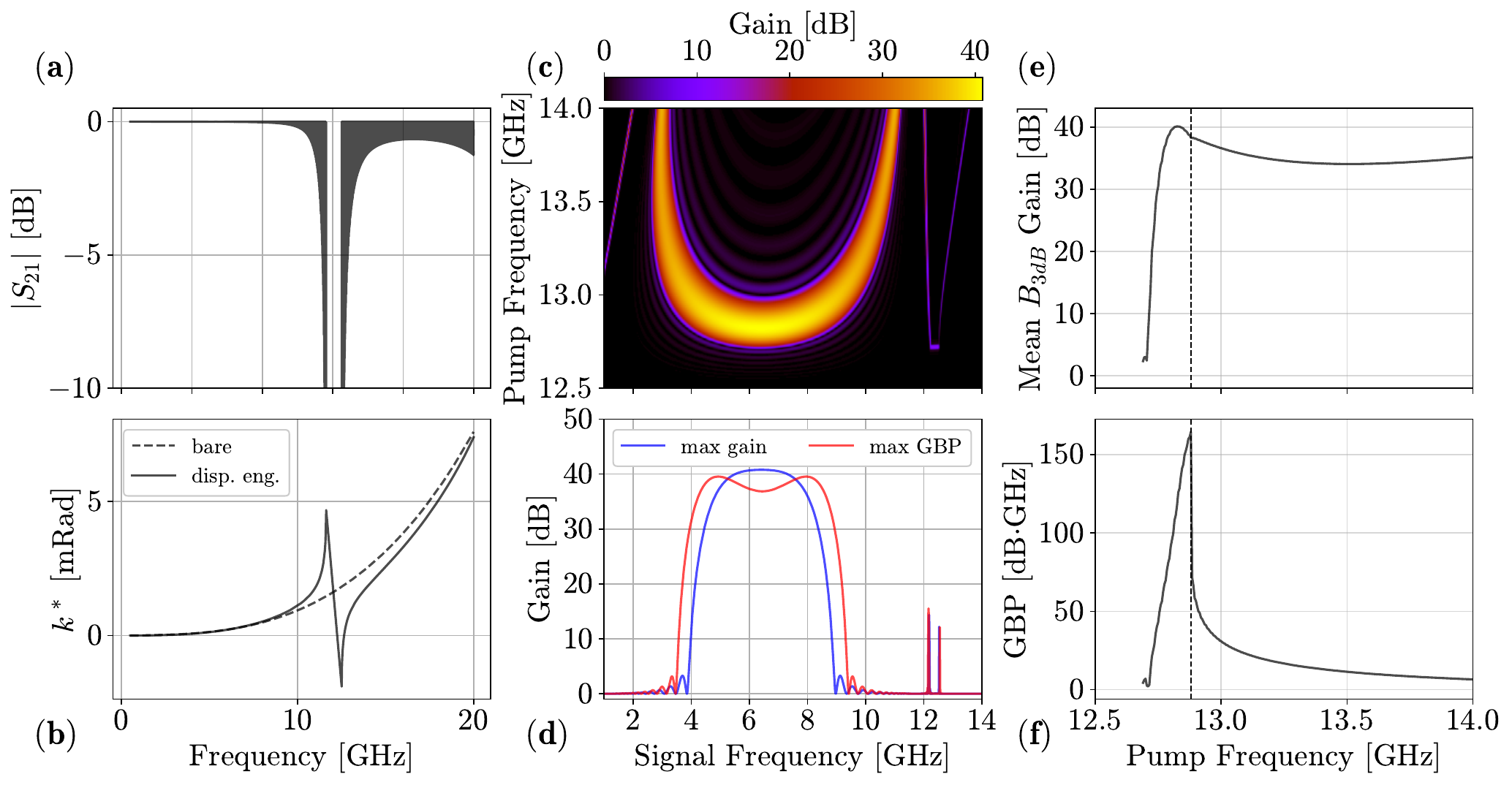}
    \caption{Simulating a full device. Parameters are given in Table~\ref{tab:imskit_v3_sim_params}. \textbf{(a), (b)} Show the photonic bandgap both in the magnitude, $|S_{21}|$, and nonlinear phase relation, $k^* = \arg(S_{21}) - \omega \sqrt{\mathcal{L}_d\mathcal{C}}$, respectively. In (b) we show the dispersion relation both for a \textit{bare} device with no dispersion engineering (all unit cells are $Z_0 = 50~\Omega$), and with engineering. \textbf{(c)} Gain simulation as a function of pump frequency. \textbf{(d)} Example gain profiles at the maximum gain $\omega_p$ and maximum GBP $\omega_p$.\textbf{(e)} The mean gain across the 3~dB bandwidth, $B_{3dB}$. The bandwidth is extracted for each pump frequency by finding the frequency at which max gain occurs, and measuring when the gain has dropped by 3~dB on both sides of the max gain point. \textbf{(f)} Gain-bandwidth-product (GBP) versus pump frequency -- where GBP is the mean \B ~gain multiplied by \B. The dashed line in (e) and (f) indicate the pump frequency yielding the maximum GBP.}
    \label{fig:dev_sim}
\end{figure*}

We numerically calculate $T_\text{KIT}^{N_\text{sc}}$ to obtain the whole KIT $S$-matrix. Notably, $S_{21}$ and $\arg(S_{21})$ are quantities critical for placing the photonic bandgap appropriately such that broadband gain (i.e. phase-matching) is possible over the target frequency range. This procedure is most frequently accomplished by adjusting the supercell electrical length via $N_u$ and $N_l$. Fig.~\ref{fig:dev_sim}(a) and (b) show this portion of the design process placing the photonic stopband near 12~GHz. Finally, the linear response obtained from the ABCD matrices is used to compute the expected gain by solving the CMEs.

These simulations and modeling are automatically performed using the custom software package \textit{twpasolver} \cite{twpasolver}, which features an extended version of the standard CME framework presented in~\cite{malnou2021three} while employing an improved, computationally efficient approach to simulating the nonlinear behavior. Furthermore, the basic CME system including only the signal, pump, and idler modes often fails to reproduce the complex mixing dynamics in the nonlinear transmission line \cite{dixon2020capturing}. To obtain more faithful predictions, we introduce the ability to automatically incorporate any number of additional modes, and to consider both 3WM or 4WM relations in the CMEs. This additional functionality allows simulation of parasitic effects such as higher pump harmonic generation. As reverse-propagating modes are also incorporated, this also enables the design of more fully-featured KITs where a second pump and frequency conversion processes may be used to realize built-in directionality \cite{ranadive2024traveling, malnou2024traveling}. Moreover, while designing a KIT with gain at the desired frequency only requires knowledge of $S_{21}$, \textit{twpasolver} also incorporates reflections at the device input and output ($S_{11}$ and $S_{22}$), as well as loss along the line. These final features enable increasingly accurate simulation of real-world devices and, by taking into account phase mismatch due to imperfect impedance matching and packaging \cite{planat2020photonic, kern2023reflection}, are able to predict phenomena such as gain ripple \cite{howe2025strongly}.

\subsection{Full Device Simulation}

Once the photonic bandgap is approximately placed at the desired frequency we perform gain simulations by re-solving the CMEs. Now we inject a nonzero $I_{p0}$ and $I_{s0}$ and extract the gain with respect to various parameters. This procedure allows for refinement of the gain profile center and bandwidth -- using gain versus pump frequency sweeps to determine optimal performance. For example, the relative strength of the loadings causes larger (negative) excursions in the nonlinear dispersion relation, $k^*$. The loading strength depends on the ratio of the number of loaded cells to unloaded cells, and on the loaded cell impedance. Tuning the loading strength affects the region over which phase-matching is satisfied and can be used to improve performance (i.e. bandwidth) at a desired frequency.

Simulation of a exemplar device while sweeping the pump is detailed in Fig.~\ref{fig:dev_sim}(c)--(d) with parameters detailed in Table~\ref{tab:imskit_v3_sim_params}. From Fig.~\ref{fig:dev_sim}(c) we can extract the 3~dB gain bandwidth, $B_{3dB}$, and the gain bandwidth product (GBP) -- defined as the mean gain across $B_{3dB}$ multiplied by $B_{3dB}$.

\begin{table}[t]
    \centering
    \caption{Parameters used for simulating a KIT exemplar. $c$ is the specific capacitance and $I_{s0}$ ($I_{p0}$) are the initial signal and pump amplitudes at the device input.}
    \begin{tabular}{l l}
        \hline\hline
        \textbf{Parameter}~~~ & \textbf{Value}  \\
        \hline
        \hline
        $L_0$ & 30~pH/$\square$ \\
        $\mathcal{C}$ & 12.3~fF/$\mu$m \\
        $c$ & 0.9~fF/$\mu$m$^2$ \\
        $w_0$, $w_1$, $s$ & $1~\mu$m \\
        $\ell_{48}$ ($\ell_{78}$) & $12~\mu$m ($3.9~\mu$m) \\
        $N_u$ & 30 \\
        $N_l$ & 4 \\
        $N_\text{sc}$ & 1200 \\
        $I_*$ & 2~mA \\
        $I_{dc}$ & $220~\mu$A \\
        $I_{s0}$ & $1.4~\mu$A \\
        $I_{p0}$ & $100~\mu$A\\
        \hline\hline
    \end{tabular}
    \label{tab:imskit_v3_sim_params}
\end{table}

\begin{figure*}[t]
    \centering
    \includegraphics[width = 0.98\textwidth]{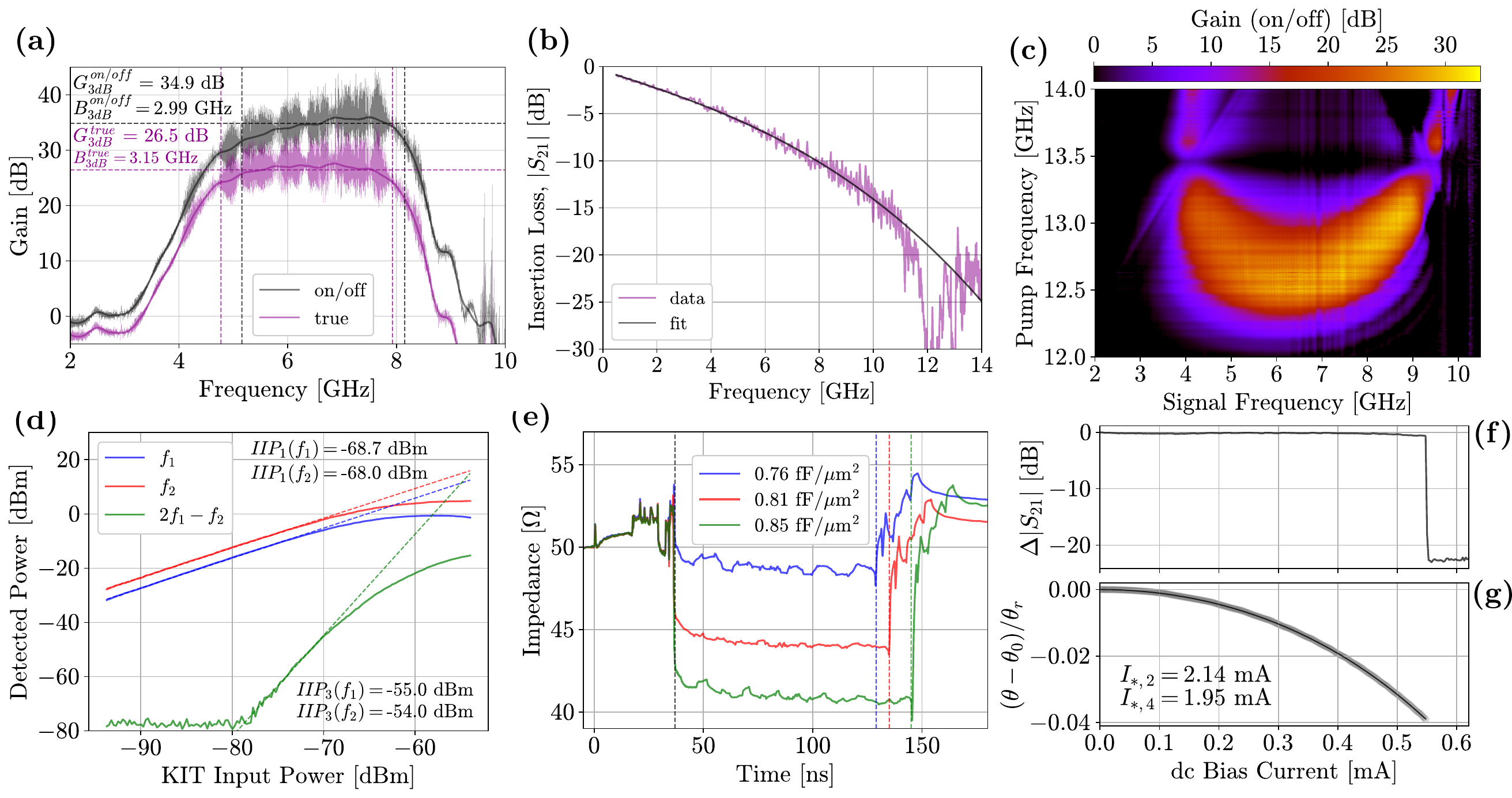}
    \caption{KIT characterization. \textbf{(a)} Gain curve corresponding to the GBP-maximized (true gain) pump frequency of 12.613~GHz after increasing the pump power to just below the onset of uncontrolled nonlinearities. The solid line shows the smoothed gain to guide the eye. The on/off gain is the ratio of the transmission with the pump on to the pump off, while the true gain is the on/off gain reduced by the KIT insertion loss. \textbf{(b)} KIT insertion loss measured at 140~mK. Ingess/egress cabling loss is removed using a cryogenic rf switch and a through-cable reference. \textbf{(c)} Pump frequency sweep gain measurement, of the same device as in (a), at fixed dc bias and pump power. \textbf{(d)} Input compression first- ($IIP_1$) and third-order ($IIP_3$) intercept points. \textbf{(e)} TDR measurements of three full devices with different stub lengths -- to compensate for differences in the dielectric's specific capacitance -- demonstrating flexible impedance targeting. Dashed colored lines indicate the start and end of the KIT. All devices have the same physical length so the difference in each device's endpoint is due solely to the unique phase velocity on each device. \textbf{(f, g)} Critical and scaling current measurements with data points shown as open circles and a line showing the fit to Eq.~(\ref{eq:theta_Istar}).}
    \label{fig:basic_testing}
\end{figure*}

\section{Experimental Results}
\begin{figure}[t]
    \centering
    \includegraphics[width=0.48\textwidth]{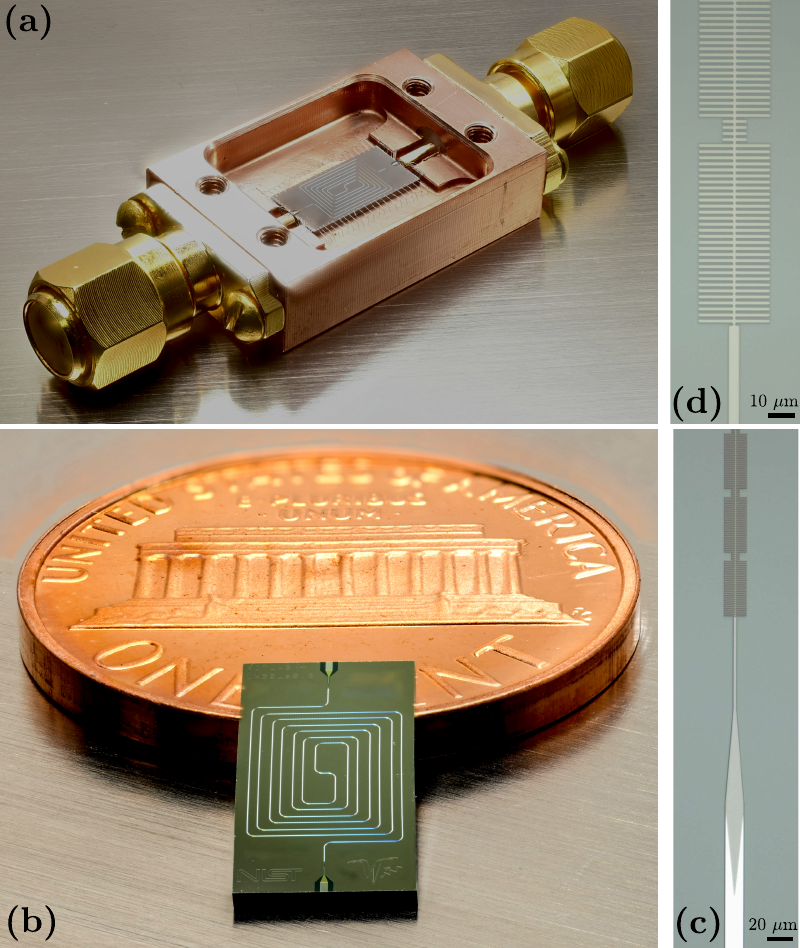}
    \caption{Fabricated KIT photos. \textbf{(a)} Device (5~mm~$\times$~10~mm die) packaged in copper enclosure (lid not shown). \textbf{(b)} Bare die. \textbf{(c)} Micrograph of the CPW-microstrip transition and first three supercells of the stub-loaded artificial TL. Photo taken immediately after patterning the NbTiN bottom metal and before deposition of the a-Si dielectric. White metal is Al, while the gray metal is NbTiN. \textbf{(d)} Higher magnification micrograph of the stub-loaded TL.}
    \label{fig:chip_pictures}
\end{figure}

Fabricated devices in this work build off of the exemplar devices simulated in the previous section with a number of design splits. To account for process variation in both the kinetic inductance and specific capacitance (from small variations in the NbTiN and a-Si thickness, respectively) we generate a number of design splits to create devices whose artificial TL is well-matched to $50~\Omega$. This is achieved by varying the finger length to best match to a given kinetic inductance and specific capacitance combination (see Sec.~\ref{sec:design_and_simulation}). Nine variants which bracket our central values of $L_0 = 30$~pH/$\square$ and $c = 0.81$~fF/$\mu$m$^2$ are created from the combinations of $L_0 = \{25, 30, 35\}$~pH/$\square$ and $c = \{0.76, 0.81, 0.85\}$~fF/$\mu$m$^2$ (corresponding to dielectric strengths of $_r = \{8.6, 9.1, 9.6\}$). Fabrication details are provided in Appendix~\ref{sec:app:fab} and techniques for extracting the kinetic inductance and specific capacitance for a given fabrication run can be found in Appendix~\ref{app:test_structures}. Photos of devices at various stages of fabrication and packaging are shown in Fig.~\ref{fig:chip_pictures}.

Devices are typically screened at 3~K in a fast-turnaround mechanically cooled cryostat using a pair of SP6T rf MEMS switches with no attenuation on the cryostat ingress/egress cabling and rf components for dc bias, pump tone, and signal probe combination located at room temperature. Devices with the best performance are identified for further testing at 50~mK in a $^3$He/$^4$He dilution refrigerator. In both systems we omit cryogenic device-level magnetic shields and use only a room temperature shield on cooldown to avoid pinning of the Earth and stray fields. Typically this shield remains installed but measurements performed with the shield removed after the cryostat is cold show no change in performance or reliability.

\subsection{Device Tuneup and Basic Characterization}

A number of auxiliary measurements are required to fully assess any given device's capabilities and performance. These include time domain reflectometry (TDR) to determine the artificial TL $Z_0$, an $I_c$ and $I_*$ measurement, pump frequency and power sweeps, and compression point measurements. Fig.~\ref{fig:basic_testing} shows a packaged device and the results of this measurement suite for characteristic devices. 

Measurement of the device gain is performed initially using the \textit{on/off} technique wherein the KIT transmission is compared with the pump on versus with the pump off. However, in the case of non-negligible insertion loss -- which is typical for nearly all superconducting TWPAs \cite{malnou2021three, ranadive2022kerr, macklin2015near, gaydamachenko2025rf, ho2012wideband, shu2021nonlinearity, vissers2016low, ranadive2024traveling, malnou2024traveling} -- measurement of the true gain is the correct metric when comparing different architecture devices. This is accomplished by measuring the KIT insertion loss using cryogenic rf switches and comparing the unpumped transmission to a through reference with unity transmission. In this way, and as part of our standard device screening procedure, we extract a given KIT device's true gain: its on/off gain reduced by its insertion loss. 

To measure $I_c$ and $I_*$ we measure the device $|S_{21}|$ over a narrow frequency span below the photonic bandgap. We ramp the dc bias current until $I_c$ is exceeded and the transmission drops significantly due to the device transitioning to the normal state. $I_*$ may also be extracted from this measurement by tracking the relative phase shift of the VNA probe and taking advantage of the fact that increasing the bias, $I_{dc}$, increases $\mathcal{L}$ and thus the electrical length of the KIT. By comparing the fractional phase shift as a function of the bias to the zero bias total phase shift, $\theta_r$, we can extract both the second- and fourth-order scaling currents for the $1 \times 0.01~\mu$m NbTiN central line \cite{shu2021nonlinearity}. This is accomplished by fitting the probe tone phase to
\begin{equation}\label{eq:theta_Istar}
    \frac{\theta - \theta_0}{\theta_r} = - \left( \frac{I_{dc}}{I_{*,2}} \right)^2 - \left( \frac{I_{dc}}{I_{*,4}} \right)^4,
\end{equation}
which follows from Eq.~(\ref{eq:kin0}) and the fact that $v_\text{ph} = 1/\sqrt{\mathcal{LC}}$. $\theta$ and $\theta_0$ are the bias-current-dependent and zero-bias probe tone phases measured by the VNA, respectively \cite{zmuidzinas2012superconducting}. $\theta_r = \omega \tau / 2$ is the reference phase -- where $\omega$ is the probe frequency and $\tau$ is the full light traversal time of the device measured via TDR. This yields $I_{*,2}$ ($I_{*,4}$) of 2.14~mA (1.95~mA) where the former is more than three times lower than for $1~\mu$m$~\times 0.02~\mu$m NbTiN studies previously \cite{malnou2021three}. If only the quadratic component of Eq.~(\ref{eq:theta_Istar}) is considered there is a clear quartic residual so we include both terms.

Measurement of the input compression points $IIP_1$ and $IIP_3$ \cite{pozar2012microwave} is accomplished by injecting two probe tones and monitoring -- on a spectrum analyzer -- the readout chain output power at the probe tone frequencies, $f_1$ and $f_2$, as well as the lowest order intermodulation distortion (IMD) product supported by 3WM: $2 f_1 - f_2$ \cite{remm2023intermodulation}. Further, we choose $f_1 = 5.5$~GHz and $f_2 = 5.2$~GHz such that all three monitored tones lie in a region of high and consistent gain. This also places all three tones below the half-pump frequency $\omega_p / 2$, which is useful for avoiding confusion between the signals, idlers, and IMD products. We extract an $IIP_1$ ($IIP_3$) of -68~dBm (-55~dBm); demonstrating more headroom between $IIP_1$ and $IIP_3$ in our 3WM KITs compared to recent 4WM JTWPA studies \cite{remm2023intermodulation}. This increased dynamic range is plausibly explained by the fact that the lowest order IMD product in 4WM mixing involves the pump, while it does not for 3WM. Since the pump is more than two orders of magnitude higher power than signals and idlers this mixing element is significantly larger. Detailed studies using KITs, which natively support both 3WM and 4WM with no modifications to the dispersion engineering, or JTWPAs can definitively confirm these relations. This would inform the most suitable mixing regime for highly multiplexed readout in which $IIP_3$ limits the multiplexing factor due to crosstalk between readout channels more often than $IIP_1$ is reached via addition of all the readout tones \cite{mates2019crosstalk}.

\begin{figure*}[t]
    \centering
    \includegraphics[width=\textwidth]{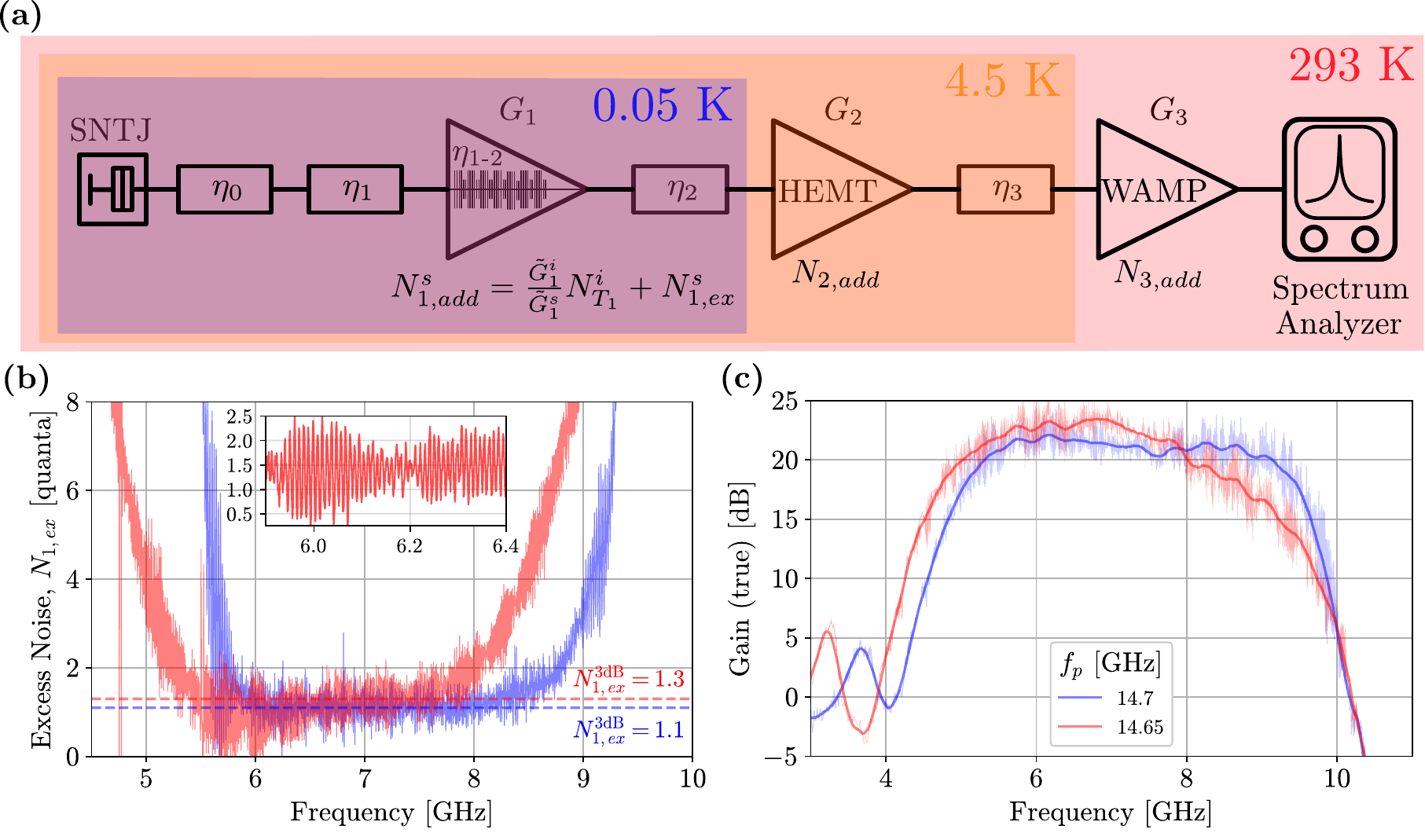}
    \caption{Measuring the system noise with a KIT FSA. \textbf{(a)} Simplified schematic of the dilution refrigerator setup used in the noise measurement. White noise is generated by an SNTJ and transmitted to the KIT input through a series of rf components with finite loss designated by $\eta_0$ and $\eta_1$. The former describes components required for the SNTJ operation, while the latter encompasses components for KIT bias and pump injection. The noise power reaching the KIT amplification medium is thus $N_\text{in} = \eta_0 \eta_1 e |V| / 2$. \textbf{(b)} System excess noise of a representative KIT at two different gain tunings showing the intrinsic limit on the KIT excess noise is approximately 1.1~quanta. The reference plane for this measurement is on the KIT die itself where amplification begins. In this case the system noise is the excess noise plus the QL of 0.5~quanta. \textbf{(c)} KIT true gain measurement demonstrating a positive detuning in the pump frequency, $f_p$, effectively compensates for the finite KIT insertion loss $\eta_{1\mathrm{-}2}$. The dark lines are smoothed curves to guide the eye.}
    \label{fig:noise_meas}
\end{figure*}

\subsection{System Noise Measurements\label{sec:noise}}

Noise is measured using a shot noise tunnel junction (SNTJ) which, under certain limiting cases, generates a calibrated white noise level dependent only on a single experimental parameter: the voltage across the junction $V$ \cite{malnou2021three, malnou2024low}. Generally, the SNTJ-generated power spectral density in photon-normalized units is
\begin{align*} 
    N_\textbf{SNTJ} = & \frac{eV + \hbar \omega}{4 \hbar \omega} \coth\left( \frac{eV + \hbar \omega}{2 k_B T_e} \right) + \\
    & \frac{eV - \hbar \omega}{4 \hbar \omega} \coth\left( \frac{eV - \hbar \omega}{2 k_B T_e} \right),
\end{align*}
with $e$ the electron charge, $\hbar$ the reduced Planck constant, $\omega$ the frequency of interest, $k_B$ the Stefan-Boltzmann constant, and $T_e$ the electron temperature. In the low-temperature, low-frequency, and high-bias limit -- i.e. $k_B T_e \ll \hbar \omega \ll e |V|$ -- the PSD simply becomes 
\begin{equation}
    N_\text{SNTJ} = \frac{e |V|}{2 \hbar \omega}.
\end{equation}

The amplifiers presented here are operated at 50~mK and provide gain typically in the range of \mbox{2--12}~GHz so $k_B T_e = 7 \times 10^{-25}$~J and $\hbar \omega =$\mbox{(1--8)$\times 10^{-24}$}~J. Therefore an SNTJ bias voltage of only $\sim 10~\mu$V is sufficient to push the noise PSD into this simple linear regime. At 5~GHz the SNTJ outputs 0.48~quanta with $V = 10~\mu$V so this regime is accessible at very low photon numbers.

A simplified experimental setup used to measure the system noise with a KIT as the FSA is shown in Fig.~\ref{fig:noise_meas}(a); while the full setup is shown in Fig.~\ref{fig:full_noise_schematic}. Here we distinguish between lossy components required for operation of the SNTJ (combined transmittivity of $\eta_0$), and those required for KIT operation ($\eta_1$). This allows us to both effectively calculate the correct white noise power incident on the KIT, but also to appropriately move the reference plane from the KIT's amplification medium input to the input of the component chain corresponding to $\eta_1$ -- i.e. where an actual device would be installed to be read out by this amplification chain \cite{malnou2024low, howe2025compact}. Transmittivity is defined as the total amount of power which propagates through a given component and includes both reflections and loss.

Note that the KIT used in the system noise measurements detailed here is a different device than that presented in Fig.~\ref{fig:basic_testing}. Further, the noise measurement KIT is housed in an upgraded package which eliminates wirebond interconnects in favor of double-ended spring probe contacts. This eliminates the large wirebond inductance at the device input and output. Subsequently this improves the KIT gain ripple via reducing impedance mismatch at the chip rf launch \cite{kern2023reflection} while alleviating parametric-oscillation-driven pump depletion to increase the maximum gain and widens the bandwidth. These performance improvements based on packaging optimizations are expected for all TWPA technologies (see Appendix~\ref{app:gripple} for further discussion). 

The SNTJ is differentially biased using a 20~Hz triangle wave from an arbitrary waveform generator (AWG) with an isolated outer shield output. The amplification chain output power is captured on a spectrum analyzer operated in zero span mode with a resolution (video) bandwidth of 1~MHz (8~MHz) and triggered using the second output of the AWG -- which is operated in pulse mode with an additional phase to ensure only the $V>0$~V portion of the triangle ramp is captured. In this way we fit only half of the shot noise curve to the asymptotic high-bias linear relation \cite{howe2025compact, malnou2024low}
\begin{equation}
    N_{3,\text{out}}^s = \tilde{G}_\text{sys}^s \left( \tilde{N}_{\text{in}}^s + \frac{\tilde{G}_1^i}{\tilde{G}_1^s} \tilde{N}_{\text{in}}^i + N_{1,\text{ex}}^s \right) + \frac{\tilde{G}_\text{sys}^s}{\tilde{G}_1^s} N_{2,\text{add}}^s.
    \label{eq:Nout3_full}
\end{equation}
The superscript $s$ ($i$) refers to the signal (idler) frequency, $\tilde{N}_\text{in}$ is the noise incident at the KIT input generated by the SNTJ, $\tilde{G}_\text{sys}$ is the system gain, and $\tilde{G}_1$ is the KIT true gain. $N_{1,\text{ex}}^s$ is the degree to which the KIT fails to meet the QL: the \textit{excess noise}. Tildes indicate values which are transformed by the finite insertion loss between components or gain stages \cite{howe2025compact, malnou2024low}. 

The formulation of Eq.~(\ref{eq:Nout3_full}) extracts the system excess noise at the input the KIT's amplification medium -- upstream of $\eta_1$ incurred via the necessity of incorporating lossy external components for bias and pump injection -- so it is correct to express the excess noise as an untransformed quantity. This correction and corresponding reduction in performance can be effectively neglected in the event the bias and injection may be accomplished using rf circuitry integrated on the KIT die using lossless superconductors \cite{howe2025compact, howe2025integrated}. In the high-KIT gain, low-HEMT-noise limit we may neglect the final term in Eq.~(\ref{eq:Nout3_full}), which is valid over the entire frequency range shown in Fig.~\ref{fig:noise_meas}(b). Outside this range of validity the primary artifact of this simplification is to falsely attribute the HEMT-added noise referred to the KIT input as KIT excess noise. Note in Eq.~(\ref{eq:Nout3_full}) we have already discarded the added noise of the warm amplifiers (WAMP) as the KIT+HEMT gain boosts the QL to approximately 5000~K.

Fig.~\ref{fig:noise_meas}(b) shows the results of the noise measurement at two different tuning points for a representative KIT; specifically a tuning which creates the optimally flat and symmetric gain profile in the on/off case ($f_p = 14.65$~GHz), and one which compensates for the KIT insertion loss to create the corresponding flat, symmetric \textit{true} gain profile ($f_p = 14.7$~GHz). The latter tuning extends $B_{3dB}$ and Fig.~\ref{fig:noise_meas}(c) shows the gain curves of each tuning point. From the system noise measurement we observe that when the KIT gain is approximately above 20~dB the system noise asymptotically reaches a minimum mean excess noise of 1.1~quanta. Further, we demonstrate that the positive pump detuning, which flattens and widens the true gain profile, also meaningfully manifests as an increase in the \textit{noise bandwidth}. I.e. the bandwidth at which the KIT operates to within a factor of two of its intrinsic limit ($N_{ex} \sim 1.1$~quanta). The noise bandwidths are 3.4~GHz and 3.1~GHz for $f_p = 14.7$~GHz and 14.65~GHz, respectively, and this demonstrates the utility of this tuning technique. Finally, in addition to being amongst the lowest KIT noise measured, this is to our knowledge the first demonstration of an nQL TWPA (JTWPA or KIT) without magnetic shielding -- underscoring the resilience of kinetic-inductance-based parametric amplifiers to magnetic fields.

\subsection{Moving the Reference Plane}
\begin{figure}
    \centering
    \includegraphics[width=0.49\textwidth]{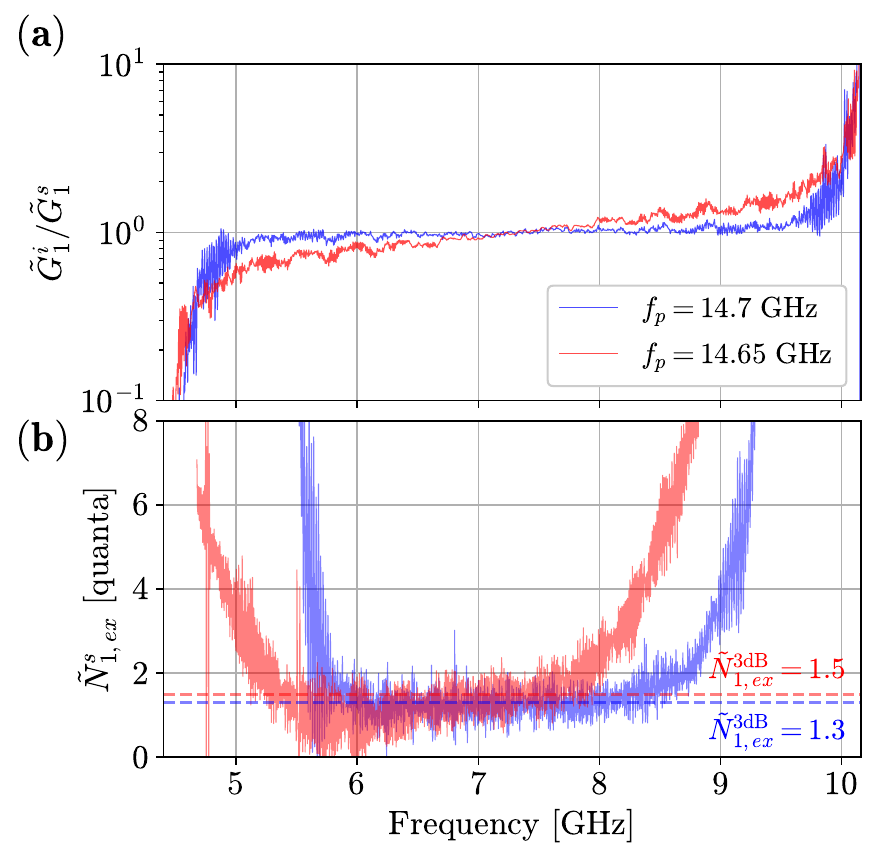}
    \caption{Moving the reference plane via transforming the excess noise. \textbf{(a)} KIT signal-idler gain asymmetry $\tilde{G}_1^i / \tilde{G}_1^s$ calculated from the raw true gain $\tilde{G}$ shown in Fig.~\ref{fig:noise_meas}(c). \textbf{(b)} Transformed system excess noise seen by a DUT located at the input of $\eta_1$. The total system noise at this reference plane is the KIT excess noise plus the vacuum (0.5~quanta).}
    \label{fig:Gi_over_Gs_Nex_tilde}
\end{figure}

The measurements of Sec.~\ref{sec:noise} correspond to the system noise at the reference plane of the amplifier input, i.e. to the right of $\eta_1$, which is on-chip and in a location inaccessible to the experimenter. To obtain the system noise experienced by a DUT at the input of the composite device composed of the KIT and its mandatory external hardware, which is how we should think of such an amplifier, we move the reference plane to be between $\eta_0$ and $\eta_1$. Thus we transform the excess noise of Fig.~\ref{fig:noise_meas}(b) according to \cite{malnou2024low}
\begin{equation}\label{eq:Nex_transform}
    \tilde{N}_{1,\text{ex}}^s = \frac{(1 - \eta_1^s) N_{T_1} + N_{1,\text{ex}}^s}{\eta_1} + \frac{\tilde{G_1^i}}{\tilde{G_1^s}}\frac{(1 - \eta_1^i)N_{T_1} + N_{1,\text{ex}}^i}{\eta_1}.
\end{equation}
$N_{T_1}$ is the thermal photon occupancy (variance) as a function of frequency at temperature $T_1 = 0.05$~K; which is 0.5~quanta at these frequencies. If we assume both that the signal-idler gain asymmetry, $\tilde{G}_1^i / \tilde{G}_1^s$, can be explicitly calculated from the true gain of Fig.~\ref{fig:noise_meas}(c), and that the excess noise of Fig.~\ref{fig:noise_meas}(b), extracted via Eq.~(\ref{eq:Nout3_full}), is evenly split between signal and idler contributions -- $N_{1,ex}^s \rightarrow N_{1,ex}^s / 2 + N_{1,ex}^i / 2$ -- we obtain the correct excess noise experienced by a DUT at the input of $\eta_1$. 

The components which comprise $\eta_1$ are a bias tee and directional coupler, whose response has been measured cryogenically to be flat over the noise measurement frequency range, with transmittivities of -0.2~dB and -0.3~dB respectively (see Appendix~\ref{app:etas}). Thus, $\eta_1 = 0.94$~(-0.5~dB) is applied Eq.~\ref{eq:Nex_transform} to move the reference plane to the $\eta_1$ input. Fig.~\ref{fig:Gi_over_Gs_Nex_tilde}(a) shows the signal-idler gain asymmetry calculated from Fig.~\ref{fig:noise_meas}(c) and Fig.~\ref{fig:Gi_over_Gs_Nex_tilde}(b) shows the transformed system noise at the $\eta_1$ input reference plane. Naturally, any systematic error in determination of $\eta_1$ translates directly to error in the reference-plane-transformed system noise.

\section{Conclusion}
We have presented a detailed simulation procedure for KIT design and demonstrate improved performance of NbTiN-based 3WM devices using an inverted microstrip transmission line topology. Specifically, we show higher maximum gain, wider bandwidths, and improved KIT and system noise relative to prior CPW-based KITs \cite{malnou2021three}. We report the first simultaneous $IIP_1$ and $IIP_3$ measurement of a KIT and show $IIP_3$ exceeds $IIP_1$ by a wider margin than the best JTWPA measurement \cite{remm2023intermodulation}. In tandem with a reduction in required pump power (via thinning of the NbTiN film from 20~nm to 10~nm) our KITs are now more suitable than ever for multiplexed readout of quantum information systems. This work also represents the first realization of KITs with sheet kinetic inductances above 30~pH/$\square$ and we demonstrate effective techniques which maintain a 50~$\Omega$ characteristic impedance despite this increased inductance.

\begin{acknowledgments}
This work is supported by the NIST Innovations in Measurement Science program, the National Aeronautics and Space Administration (NASA) under Grant No. NNH18ZDA001N-APRA, the Department of Energy (DOE) Accelerator and Detector Research Program under Grant No. 89243020SSC000058, and DARTWARS, a project funded by the European Union’s H2020-MSCA under Grant No. 101027746. A.~Giachero and A.~Nucciotti are also supported by the Italian National Quantum Science and Technology Institute through the PNRR MUR Project under Grant PE0000023-NQSTI. P.~Campana is supported by the Italian Research Center on High Performance Computing, Big Data and Quantum Computing through the PNRR MUR Project under Grant  CN00000013-ICSC.

We thank M.~Malnou and J.~Aumentado of the NIST Advanced Microwave Photonics group for providing the SNTJ used for noise measurements.
\end{acknowledgments}

\bibliography{references}


\clearpage
\appendix

\renewcommand{\thefigure}{S.\arabic{figure}}
\renewcommand{\theequation}{S.\arabic{equation}}
\setcounter{figure}{0}
\setcounter{equation}{0}

\begin{titlepage}
  \centering
  \vskip 60pt
  \LARGE Near-quantum-limited Kinetic Inductance Traveling Wave Parametric Amplifiers: Methodology and Characterization -- Supplementary Information \par
  \vspace{5mm}
\end{titlepage}

\section{Model for extracting the dispersion-engineered transmission line parameters}\label{sec:app:model}
For any transmission line the characteristic impedance is given by $Z_0 = \sqrt{\mathcal{L} / \mathcal{C}}$ where $\mathcal{L}$ and $\mathcal{C}$ are the inductance and capacitance per unit length, respectively. The TWPA amplifier presented in this work is implemented as a stub-loaded inverted-microstrip artificial transmission line \cite{bahl2003lumped,martín2015artificial}. We form the transmission line via a NbTiN wiring layer deposited onto a high-resistivity silicon substrate, covered by an amorphous silicon (\aSi) dielectric layer, finally topped by a Nb sky (ground) plane~\cite{giachero2024kinetic,howe2025compact,shu2021nonlinearity}. The elementary cells of the transmission line have a length of $2s$ and are composed of a series inductance flanked by two stubs capacitively coupled to ground (Fig~\ref{fig:ims}). $\mathcal{L}$ is defined via the geometric and NbTiN film kinetic inductance (which is dominant), while $\mathcal{C}$ has a simple parallel plate capacitor dependence defined by the stub geometry (width $w_1$ and length $\ell$) and the a-Si permittivity, $\epsilon_r$, and thickness, $d$. By varying $\ell$ we vary $\mathcal{C}$ and, consequently, $Z_0$. The estimation of the stub length that provides the desired $Z_0$ is performed using electromagnetic simulations to extract $\mathcal{L}$ and $\mathcal{C}$ in the dc limit ($\omega \to 0$). Below we detail the simulations used for extracting the correct stub lengths to target a specific $Z_0$ for the artificial transmission line.

\begin{figure}[t]
    \centering\includegraphics[width=0.48\textwidth]{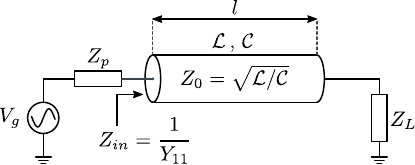}
    \caption{Transmission line connected to a generator and a load. With a 1-port simulation we determine $\mathcal{L}$ ($\mathcal{C}$) by setting the load impedance to $Z_L$ to zero (infinity). $V_g$ is the microwave generator that generates the stimulus signal at the input port during simulation, and $Z_p$ is the impedance of the port, generally set to 50~$\Omega$. \label{fig:TL} } 
\end{figure}

Considering the model shown in Fig.~\ref{fig:TL}, the input port impedance $Z_{in}$ for a transmission line with characteristic impedance $Z_0$, physical length $l$, capacitance and inductance per unit length $\mathcal{C}$ and $\mathcal{L}$, and terminated with an arbitrary load impedance $Z_L$, can be written as \cite{pozar2012microwave}: 

\begin{equation}
    Z_{in}(l)=Z_{11}(l)=Z_0 \cfrac{Z_L+Z_0\tan{(\gamma l)}}{Z_0+Z_L\tan{(\gamma l)}}.
\end{equation}
$\gamma=\alpha+j\beta$ is the propagation constant. The real part of the propagation constant, $\alpha$, is the attenuation constant while the imaginary part $\beta=\omega\sqrt{\mathcal{L}\mathcal{C}}$ is the phase constant. In case of an ideal, lossless superconducting line ($\alpha=0$) composed of $n$ unit cells which can be considered in the lumped element regime, the input port impedance can be written as
 \begin{equation}
   Z_{in}(n)=Z_{11}(n)=Z_0 \cfrac{Z_L+j Z_0\tan{(\beta n)}}{Z_0+j Z_L\tan{(\beta n)}},
\end{equation}
where $\mathcal{C}$ and $\mathcal{L}$ are taken in the quasi-static condition ($\omega\to 0$), and $n$ is the number of elementary cells, of width $2s$, that compose the transmission line.

\begin{table}[t] 
\centering
\caption{List of variables used to model the artificial transmission line.}
\begin{tabular}{cl}
\hline
\hline
\multicolumn{1}{l}{\textbf{Variable}} & \textbf{Variable} \\
\multicolumn{1}{l}{\textbf{name}} & \textbf{definition} \\ \hline\hline
$Z_0$ & line characteristic impedance\\
\hline
$L$     & line inductance per unit length/cell\\
$L_1$   & line inductance at the Port-1\\
$\mathcal{L}$  & line inductance per unit length/cell for $\omega \to 0$\\
\hline
$C$   & line capacitance per unit length/cell\\
$C_1$   & line capacitance at the Port-1\\
$\mathcal{C}$   & line capacitance per unit length/cell for $\omega \to 0$\\
\hline
$Z_{11} $ & input $Z$-parameter, \\
           &in a single-port network $Z_{in}=Z_{11}$\\ 
$Y_{11} $ & input $Y$-parameter, \\
           &in a single-port network $Z_{11}=1/Y_{11}$\\ 
\hline
$Z_{L} $ & line load impedance \\
$Z_{p} $ & port impedance, usually $50~\Omega$ \\
\hline\hline
\end{tabular} 
\label{tab:variables}
\end{table}

In the case of a short-circuit termination ($Z_L\to 0$) the line input impedance is given by
\begin{equation}
    Z_{in}=j Z_0\tan{(\beta n)}=
    j \sqrt{\cfrac{\mathcal{L}}{\mathcal{C}}} 
    \tan{\left(\omega\sqrt{\mathcal{L} \mathcal{C}}\right)}, 
\end{equation}
and describes a purely reactive impedance. Under the condition $\beta n \ll 1 \Leftrightarrow \omega\to 0$ all the current passes through the short producing a strong magnetic field with stored magnetic energy and the transmission line acts as a pure inductor. In this case $Z_{in}$ can Taylor-expanded as

\begin{equation}
    j\omega L_1 \underset{\substack{\uparrow \\[0.4em] 
        {\omega\to 0}}}{\simeq}
    j\omega\mathcal{L} n\left[1+
    \cfrac{\omega^2 n^2}{3} \mathcal{L} \mathcal{C}+
    \cfrac{\omega^4 n^4}{15} \mathcal{L}^2 \mathcal{C}^2
    \right]+\mathcal{O}(\omega^6).
\end{equation}
Thus the line input inductance $L_1$ can be expressed as:

\begin{equation}\label{eqapp:fitL}
L_1\simeq
\mathcal{L} n\left[1+
\cfrac{\omega^2 n^2}{3} \mathcal{L} \mathcal{C}+
    \cfrac{\omega^4 n^4}{15} \mathcal{L}^2 \mathcal{C}^2 
\right]
\end{equation}
 The line input inductance for the entire line is simply $L_1(\omega)=n L(\omega)$, where $L(\omega)$ is the inductance per cell. To first order, and when $\omega \to 0$, the impedance tends to its zero-frequency value, $L \to \mathcal{L}$. For $\omega \neq 0$, $L_1$ can be modeled as an effective inductance that depends on both the inductance and capacitance per cell. 
 
 Performing a 1-port simulation at low frequencies ($\omega/2\pi<100$~MHz) with a line composed of $n$ elementary cells, and with $Z_L = 0$, we obtain the input admittance $Y_{11}$ (where $Z_{11} = 1/Y_{11}$ in the case of single-port network) as function of the stub length $\ell$. At low frequency, a shorted line can me modeled as a simple inductive element with reactance $j\omega L_1=\operatorname{Im}\left(Z_{11}\right)$. From $Y_{11}$ we obtain the line input inductance as $L_1=\operatorname{Im}\left(1/Y_{11}\right)/\omega$. Fitting the $L_1$ trend to Eq.~(\ref{eqapp:fitL}) yields $\mathcal{L}$ and $\mathcal{C}$ as a function of $\ell$.   

Similarly, for an open termination ($Z_L\to \infty$) the line input impedance is given by
\begin{equation}
    Z_{in}=-j Z_0\cot{(\beta n)}=
    -j \sqrt{\cfrac{\mathcal{L}}{\mathcal{C}}} 
    \cot{\left(\omega\sqrt{\mathcal{L} \mathcal{C}}\right)},
\end{equation}
and once again describes a purely reactive impedance. When $\beta n \ll 1 \Leftrightarrow \omega\to 0$ there is no current passing through the central conductor and the transmission line acts as pure capacitance. Now $Z_{in}$ is Taylor-expanded as

\begin{equation}
-\cfrac{1}{j\omega C_1}  \underset{\substack{\uparrow \\[0.4em] 
        {\omega\to 0}}}{\simeq}
    -\cfrac{1}{j\omega\mathcal{C} n}\left[1-
    \cfrac{\omega^2 n^2}{3} \mathcal{L} \mathcal{C}-
    \cfrac{\omega^4 n^4}{45} \mathcal{L}^2 \mathcal{C}^2
    \right]+\mathcal{O}(\omega^5).
\end{equation}
The capacitance, $C_1$, seen at the line input is thus

\begin{equation}\label{eqapp:fitC}
C_1\simeq
\mathcal{C} n\left[
\cfrac{45}{45-15 \omega^2 n^2 \mathcal{L} \mathcal{C}-\omega^4 n^4 \mathcal{L}^2 \mathcal{C}^2}
\right].
\end{equation}

This is the input line capacitance for the entire line, which can be expressed as $C_1(\omega) = n C(\omega)$, where $C(\omega)$ is the capacitance per cell. With $\omega\to 0$ the capacitance per cell tends to its zero-frequency value $C\to \mathcal{C}$. For higher frequency $C_1$ instead depends on both the inductance and capacitance per cell. The open line in this case is simple a capacitive element with reactance $-j/(\omega C_1)=\operatorname{Im}\left(Z_{11}\right)$ and the input capacitance is given by $C_1=-1/[\omega\cdot\operatorname{Im}\left(1/Y_{11}\right)]$. By fitting the obtained $C_1$ trend to Eq.~(\ref{eqapp:fitC}) we obtain $\mathcal{L}$ and $\mathcal{C}$ as a function of $\ell$.

Both methods can be used independently to extract both $\mathcal{L}$ and $\mathcal{C}$ and the results are compatible. In the case of this work, a combination of the two methods has been used. Eq.~(\ref{eqapp:fitC}) was used to extract $\mathcal{C}$, while Eq.~(\ref{eqapp:fitL}) was used for $\mathcal{L}$. In this approach $\mathcal{L}$ and $\mathcal{C}$ can be obtained as ($\lim_{ \omega\to 0}L_1)/n$ and ($\lim_{ \omega\to 0}C_1)/n$, respectively. By combining the obtained values we obtain the $Z_0=\sqrt{\mathcal{L}/\mathcal{C}}$ vs. $\ell$ curve that can be used to select the appropriate stub length to achieve the desired characteristic impedance. 

\begin{figure}[!h]
    \centering \includegraphics[width=0.98\linewidth,clip]{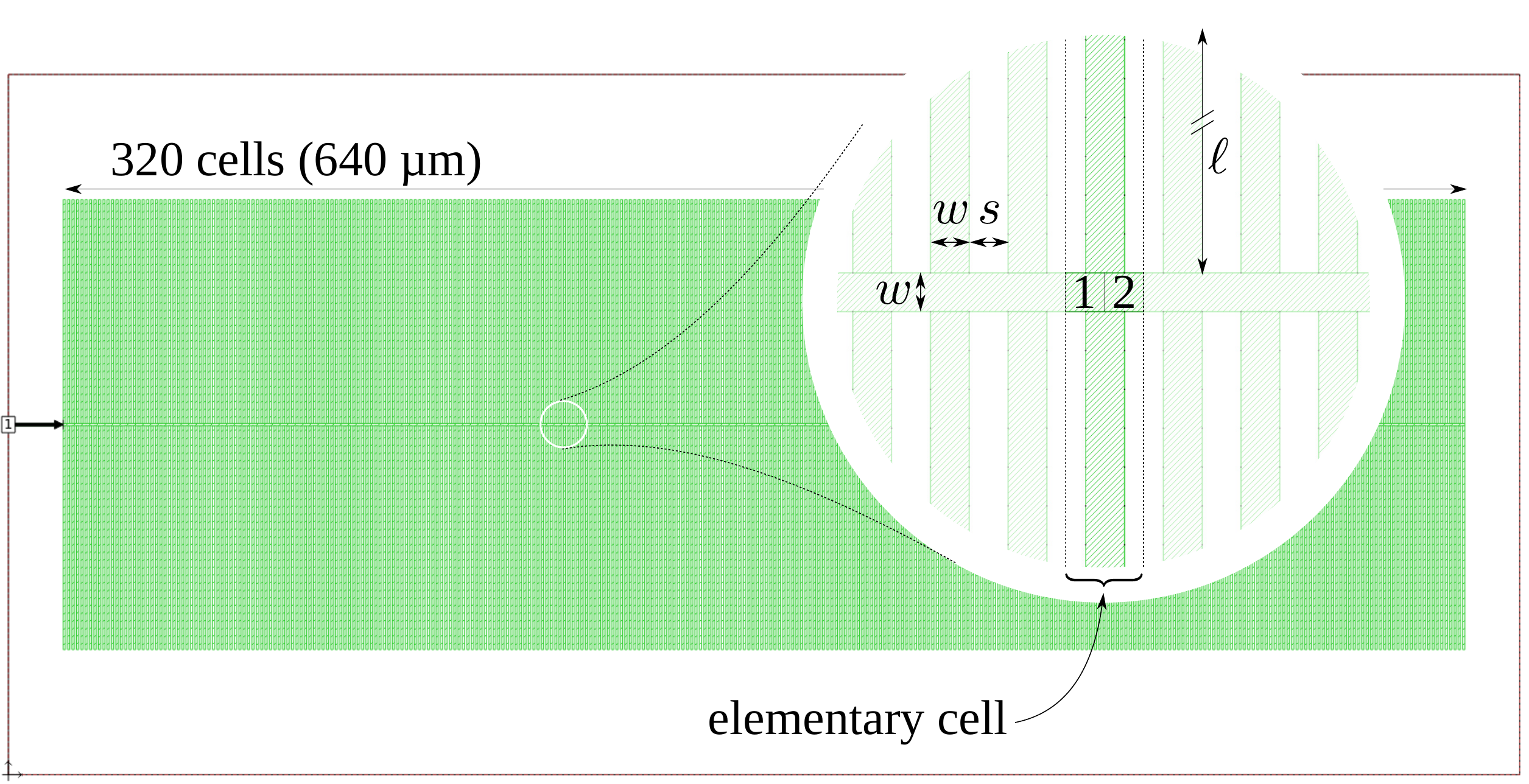}    
    \caption{Setup of the stub-loaded inverted microstrip line simulated using a commercial EM simulator. The line consists of a central conductor $640~\mu$m in length with a width of $w=1~\mu$m. Stubs, each with a width of $w=1,\mu\text{m}$, are positioned on both sides of the line at a spacing of $s=1~\mu$m. Each elementary cell includes one stub on each side and, considering the spacing, results in a total length of $2~\mu$m per cell (2~squares for the central line).} 
    \label{fig:scketup}
\end{figure}

An example of the simulated line is reported in Fig.~\ref{fig:scketup}. The line is composed of $n=320$ elementary cells (640~$\mu$m long) with stubs on both side of length $\ell \in [1,80]~\mu$m. The central conductor and the stubs have a width of $w=1~\mu$m  while the spacing between adjacent stubs is $s=1~\mu$m. Each elementary cell has a length of $2~\mu$m (2~squares for the central line). The large number of cells was chosen to replicate a realistic transmission line and which reproduces the self-interaction between stubs which are closely packed. The dielectric $\varepsilon_r=9.1$ corresponds to measured values for our a-Si and we repeat the simulations for various thicknesses $d = [100, 200, 300, 400]$~nm. The sky ground plane has negligible kinetic inductance.

\begin{figure}[!t]
    \centering \includegraphics[width=0.98\linewidth,clip]{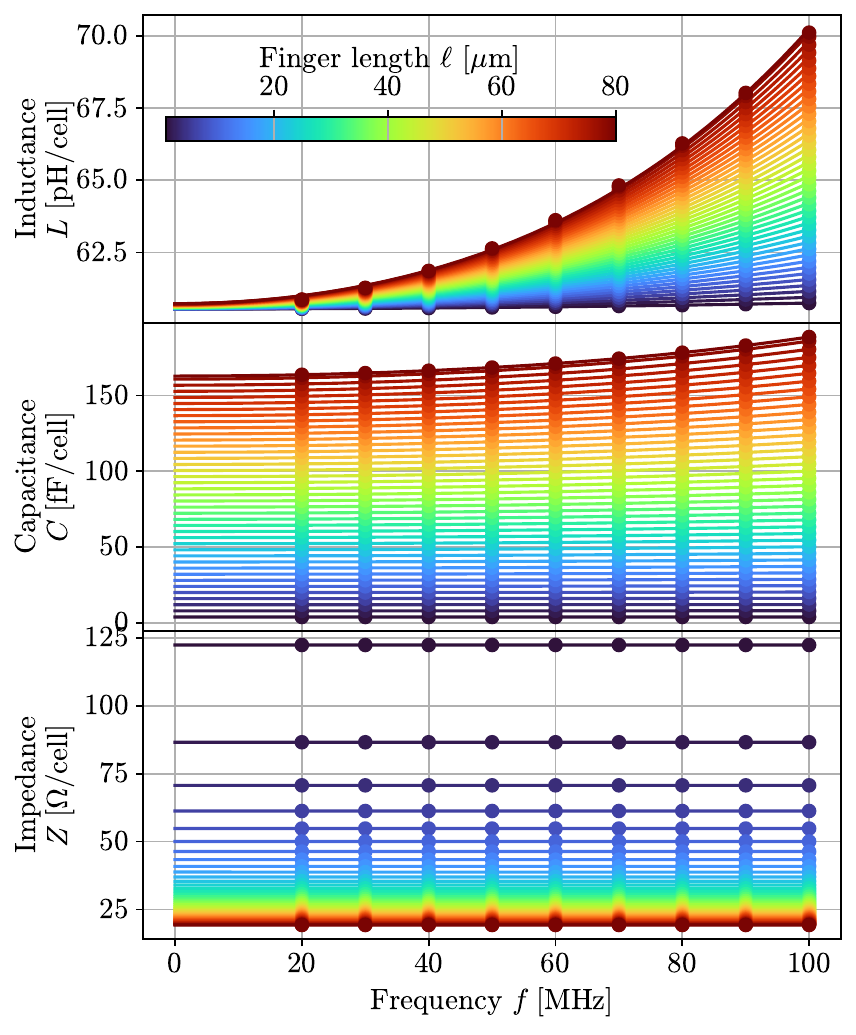}    
    \caption{Simulated inductance, capacitance and impedance per cell as a function of the frequency and on the stub length (heat-map) and with kinetic inductance $L_0=30$~pH/$\square$, dielectric permittivity $\epsilon_r = 9.1$, and dielectric thickness $d=100$~nm.}
    \label{fig:Xvsell}
\end{figure}

\begin{figure}[!t]
    \centering \includegraphics[width=0.98\linewidth,clip]{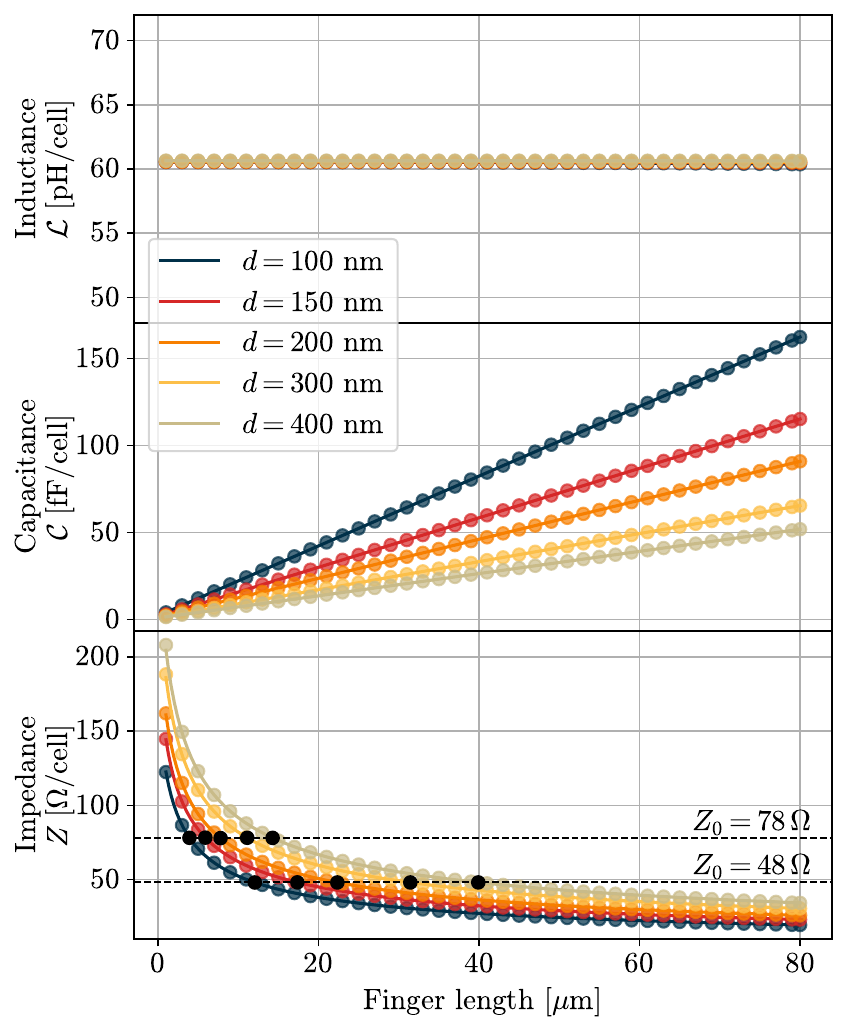}    
    \caption{Inductance, capacitance and characteristic impedance per cell as a function of the stub length at zero frequency for different dielectric thickness and with kinetic inductance $L_0=30$~pH/$\square$ and dielectric permittivity $\epsilon_r = 9.1$.}
    \label{fig:ellvsd}
\end{figure}
  
Examples of simulations and result analysis are shown in Fig.~\ref{fig:Xvsell} and Fig~\ref{fig:ellvsd}. In Fig.~\ref{fig:Xvsell} the quantities $C$ and $L$ represent $C_1$ and $L_1$ divided by the number of cells, $n$. Fig.~\ref{fig:ellvsd} shows the values of $\mathcal{L}$ and $\mathcal{C}$ estimated by the $L$ and $C$ fits and the resulting line impedance $Z_0=\sqrt{\mathcal{L}/\mathcal{C}}$. For every configuration the inductance per cell (which contains two squares) we obtain is $60.6$~pH. This indicates $\mathcal{L}$ is independent of the dielectric characteristics and depends solely on the number of squares composing the central line, as expected. Similar simulations, assuming the same kinetic inductance of $30$~pH/$\square$, a stub width of $1~\mu$m, and stub spacings of $2~\mu$m and $3~\mu$m, resulted in line inductance of $90$~pH and $120$~pH, respectively. The capacitance per cell increases linearly with the stub length and this increase rises as the dielectric thickness decreases. This behavior validates the assumption that the unit cell capacitance is strongly in the parallel-plate capacitor regime. 

Fig.~\ref{fig:ellvsd} and Table~\ref{tab:ell} report the stub lengths needed for obtaining impedances of $Z_0^u=48~\Omega$ and $Z_0^l=78~\Omega$. These impedance values are selected for creating the unloaded ($u$) and loaded ($l$) supercell sections that compose the amplifier line (see Sec.~\ref{sec:design_and_simulation}). Considering a supercell composed of $N_u$ unloaded cells and $N_l$ loaded cells the effective value of $L^{sc}$ and $C^{sc}$ can be written as:
\begin{equation}
    L^{sc}=\cfrac{N_u \mathcal{L}_u+
                                    N_l \mathcal{L}_l}
                                   {N_u+N_l}
    \quad,\quad
    C^{sc}=\cfrac{N_u \mathcal{C}_u+
                                    N_l \mathcal{C}_l}
                                   {N_u+N_l}.    
\end{equation}
Thus the supercell effective impedance is
\begin{equation}
    Z^{sc}_0=\sqrt{\cfrac{L_{sc}}{C_{sc}}}
                              =\sqrt{\cfrac{
                                            N_u \mathcal{L}_u+
                                            N_l \mathcal{L}_l}
                                           {N_u \mathcal{C}_u+
                                            N_l \mathcal{C}_l}}.
\end{equation}
In the case of $N_u=30$ and $N_l=4$ the values of $(L_u$, $C_u)$ and $(L_l$, $C_l)$ yielding elementary cell impedances of $Z^u_0=48~\Omega$ and $Z^l_0=78~\Omega$, generate a supercell (and thus full amplifier) impedance close to $50~ \Omega$.

\begin{table}[h]
\caption{Simulation results for the stub length needed for $48~\Omega$ and $78~\Omega$ elementary cells as a function of the dielectric thickness. Simulations performed with kinetic inductance $L_k=30$~pH/$\square$ and dielectric permittivity $\epsilon_r = 9.1$.
}
\begin{ruledtabular}
\begin{tabular}{llllll}
$d$ & $\ell$ & $\mathcal{L}$ & $\mathcal{C}$ &  $Z_0$ &  $Z^{sc}_0$\\\relax
[nm] & [$\mu$m] & [pH] & [fF] &  [$\Omega$] &  [$\Omega$]\\ \hline
\multirow{2}{*}{100} &          3.9 &     60.6 &     9.9 & 78.0 &\multirow{2}{*}{49.9}  \\       
    &         12.1 &     60.6 &     26.3 & 48.0  \\ \hline   
\multirow{2}{*}{150} &          5.9 &     60.6 &     10.0 & 78.0 &\multirow{2}{*}{49.9}  \\       
    &         17.4 &     60.6 &     26.3 & 48.0  \\ \hline   
\multirow{2}{*}{200} &          7.8 &     60.6 &     10.0 & 78.0 &\multirow{2}{*}{49.9} \\     
    &         22.4 &     60.6 &     26.3 & 48.0  \\ \hline   
\multirow{2}{*}{300} &         11.2 &     60.7 &     10.0 & 78.0 &\multirow{2}{*}{49.9} \\    
    &         31.5 &     60.7 &     26.3 & 48.0  \\  \hline   
\multirow{2}{*}{400} &         14.3 &     60.7 &     10.0 & 78.1 &\multirow{2}{*}{49.8} \\     
    &         40.0 &     60.7 &     26.4 & 48.0  \\
\end{tabular}
\end{ruledtabular}

\label{tab:ell}
\end{table}

For the amplifier presented here, a dielectric thickness of $d=100$~nm was selected to achieve shorter finger lengths for impedance matching the line -- enabling a more compact amplifier design.

\begin{figure}[t]
    \centering
    \includegraphics[width = 0.48\textwidth]{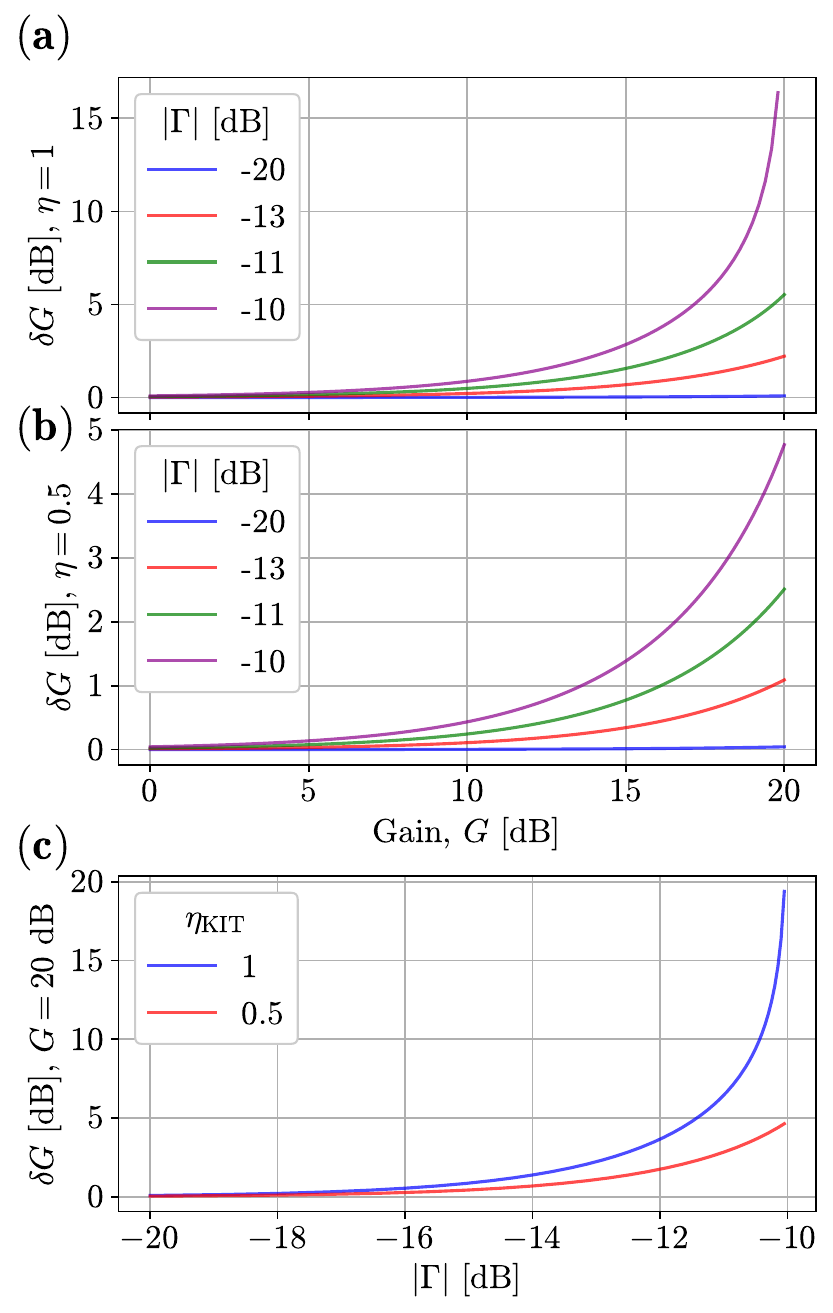}
    \caption{Analytic gain ripple estimation. \textbf(a) Gain ripple as a function of gain assuming a lossless device and for various $|\Gamma|$. \textbf{(b)} Same as (a) but for a device with $\eta_\text{KIT} = 3$~dB of total insertion loss. \textbf{(c)} Gain ripple assuming 20~dB of max gain as a function of $|\Gamma|$ for a lossless device and a device with 3~dB of insertion loss.}
    \label{fig:gripple}
\end{figure}

\section{Analytic Estimation of Gain Ripple from Impedance Mismatch \label{app:gripple}}
Gain ripple in TWPAs arises from either phase mismatch or standing wave interference along the device TL. We can analytically approximate the amount of peak-to-peak gain ripple, $\delta G$, by considering a single mode which experiences a reflection at each end of the TL. Assuming the reflections, $\Gamma$, are symmetric and the mode experiences a total phase shift upon returning to the TWPA input -- including both the phase shift from the reflection and the phase winding by traveling the full length of the device -- of $\pi$, then the gain ripple in dB is given by
\begin{equation}\label{eq:gripple}
    \delta G = 10 \operatorname{log}_{10} \left( \frac{1 + G \eta |\Gamma|^2}{1 - G \eta |\Gamma|^2} \right).
\end{equation}
$G$ is the total TWPA gain and $\eta$ is the TL transmittivity which, in this case, is equal to the insertion loss as the reflective component has been considered separately via $\Gamma$. Clearly Eq.~(\ref{eq:gripple}) diverges when $G \eta |\Gamma|^2 = 1$ -- i.e $\delta G \rightarrow G$. We can also see that gain ripple becomes increasingly significant the higher the device's maximum gain, but also that finite insertion loss serves to alleviate gain ripple. Thus, when considering the gain ripple in an imperfectly matched TWPA, loss is a feature, not a bug. Fig~\ref{fig:gripple} shows various results of these calculations and underscores the importance of improving packaging in realizing TWPAs with ever increasing performance.

\section{Device Fabrication}\label{sec:app:fab}
\begin{figure}[!h]
    \centering\includegraphics[width=1\linewidth]{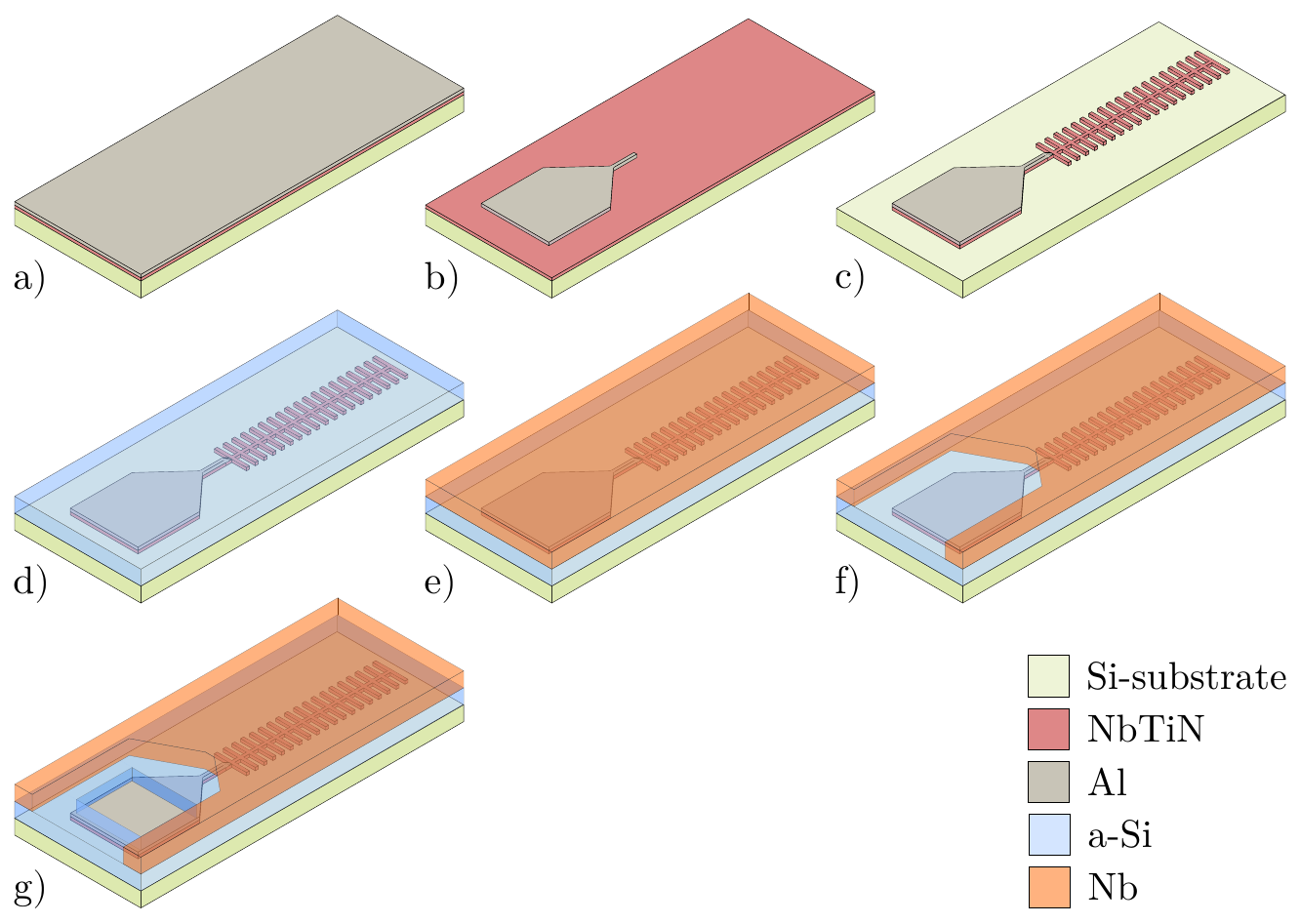}
    \caption{Fabrication flow for inverted microstrip transmission line. \textbf{(a)} 10~nm NbTiN + 50~nm Al film growth. \textbf{(b)} Al pattern and etch. \textbf{(c)} NbTiN pattern and etch. \textbf{(d)} 100~nm a-Si dielectric deposition. \textbf{(e)} 200~nm Nb sky plane deposition. \textbf{(f)} Sky plane pattern and etch. \textbf{(g)} a-Si via etch to expose bottom layer bondpad.}
    \label{fig:fab} 
\end{figure}

The IMS KITs  are fabricated on 76.2~mm, high-resistivity, intrinsically-doped, float-zone-Si wafers. Immediately before loading into the vacuum chamber, the native oxide on the Si is etched away using HF. As with previous designs~\cite{giachero2024kinetic}, the NbTiN is reactively co-sputtered from Nb and Ti targets in an Ar:N$_2$ atmosphere with the substrate held at 500~C. The fluxes of the Nb and Ti targets are tuned to maximize the film $T_c$. After the 10~nm NbTiN growth, the substrate is cooled to room temperature and 50~nm of Al is deposited in situ.  Since the two superconducting layers are in pristine electrical contact, they form a low kinetic inductance but high $T_c$ bilayer. The wafer is patterned using standard photolithography techniques using an i-line stepper and the process flow is shown in Fig.~\ref{fig:fab}.

First the Al is patterned and etched using Transene-A wet-etchant, leaving Al in the bondpad area for a later stop etch and a low inductance region for the CPW-microstrip transition.  The Al etchant does not affect the NbTiN beneath. Then the NbTiN transmission line is patterned and etched in a CHF$_3$ based ICP-RIE. This etch slowly etches the Si substrate, to ensure good coverage in later deposited layers. Next, 100~nm of insulating a-Si is deposited at room temperature in an ICP-PECVD, followed by 100~nm of sputtered Nb. The Nb ground plane is patterned and etched in a CF$_4$ based RIE which is selective against the a-Si. This forms the CPW launch for bonding the device into the package. Finally, vias in the a-Si to contact the bondpads are patterned and etched in an SF$_6$ based RIE which stops on the Al protecting the thin NbTiN layer beneath.

\section{Process Control Test Structures: Directly Measuring $L_0$ and $\epsilon_r$ \label{app:test_structures}}
As detailed in Sec.~\ref{sec:design_and_simulation}, the most critical parameters for designing IMS KITs with a  50~$\Omega$ characteristic impedance and gain located at the desired frequencies are the NbTiN $L_0$ and the a-Si dielectric strength $\epsilon_r$. These in turn determine $\mathcal{L}$ and $\mathcal{C}$ and thus directly impact $Z_0 = \sqrt{\mathcal{L} / \mathcal{C}}$ and the photonic bandgap frequency (via altering the phase velocity $v_{ph} = 1/\sqrt{\mathcal{LC}}$ and the supercell electircal length). Deviations in these parameters in fabricated devices from those assumed during design thus result in sub-optimally-matched devices and may miss the target gain frequency.

To combat these effects in subsequent iterations we co-fabricate process control test structures for measurement of $L_0$ and $\epsilon_r$. These test structures consist of (a) a dc test structure made from a straight wire of 500 squares of NbTiN (1~$\mu$m~$\times~500~\mu$m and 2~$\mu$m~$\times~1000~\mu$m), (b) two lumped-element microwave resonators composed of a straight segment of NbTiN and two parallel plate capacitors with a-Si as the  dielectric, and (c) two of the same structure as (b) but with dc bias lines allows for in-situ tuning of the NbTiN inductance -- and thus the resonator frequency. This permits an independent cross-validation measurement of $L_0$ at rf frequencies. In general we find good agreement between $L_0$ extracted from the dc and resonator structures.

\begin{figure}
    \centering
    \includegraphics[width=0.48\textwidth]{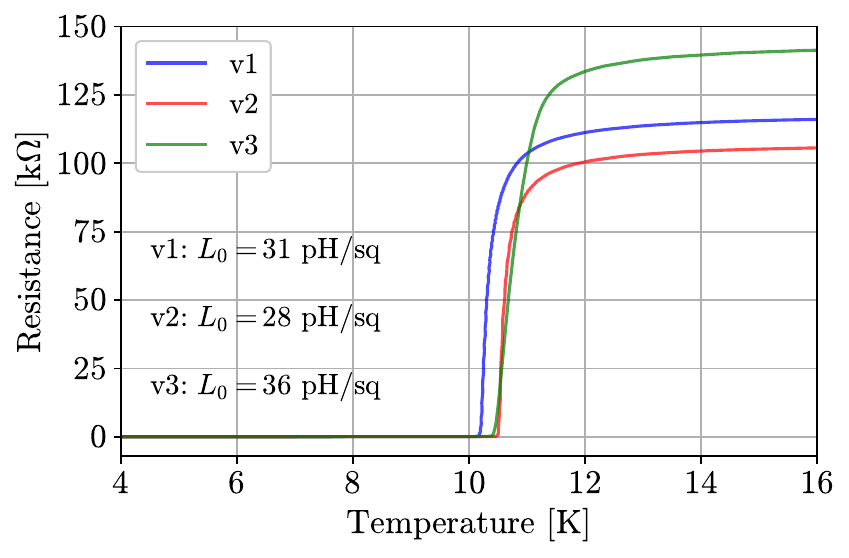}
    \caption{Zero-frequency (dc) kinetic inductance measurement of the three IMS KIT version's test structures.}
    \label{fig:Lk_dc_meas}
\end{figure}

We first determine $L_0$ using via the dc test structure by measuring its resistance as a function of temperature. The kinetic inductance of a superconducting film at zero temperature is approximately
\begin{equation}
    L_0 = \frac{\hbar R_n}{1.76 \pi k_B T_c},
\end{equation}
where $\hbar$ and $k_B$ are the reduced Planck constant and Boltzmann constants, and $R_n$ and $T_c$ are the film's normal resistance and superconducting critical temperature. By measuring the dc resistance of the test structures while the mainplate of the 3~K test cryostat is heated above 15~K we extract $R_n$ from the asymptotic resistance and $T_c$ as the midpoint of the superconducting transition. For the three fabrication runs (versions) of IMS KITs up to this point we show the results of these measurements on the 1~$\mu$m~$\times~500~\mu$m structure in Fig.~\ref{fig:Lk_dc_meas}. Both the 1~$\mu$m~$\times~500~\mu$m and 2~$\mu$m~$\times~1000~\mu$m yield self-consistent results, as expected.

\begin{figure}[!t]
    \centering
    \includegraphics[width=0.48\textwidth]{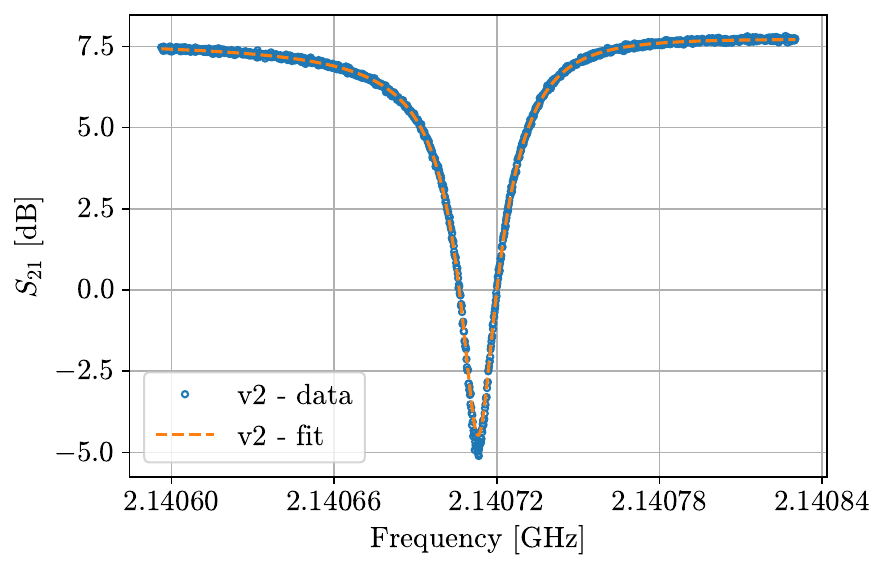}
    \caption{Example of the acquired resonance for resonator \#3 along with its fit. Each  resonance is fitted using a Python framework developed for microwave kinetic inductance detectors (MKIDs) for mm/submm astronomy~\cite{Wheeler2022}.}
    \label{fig:resonance}

    \centering
    \includegraphics[width=0.48\textwidth]{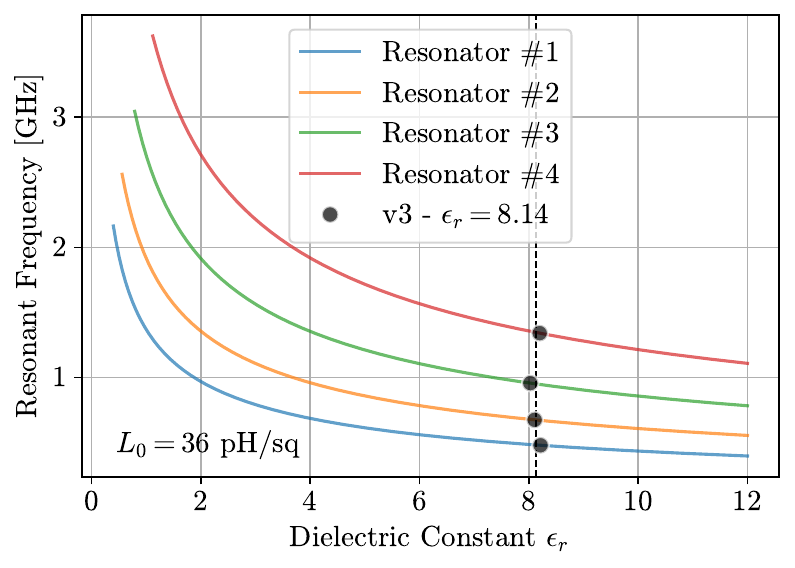}
    \caption{Example of calibration curves for the simulated resonant frequencies as a function of the of relative permittivity for a given value and kinetic inductance, compared with the measured values for the parallel plate different resonators. The vertical dashed black line shows the relative permittivity extrapolated for v3 devices.}
    \label{fig:fres_vs_eps}
\end{figure}

\begin{figure}[t]
    \centering
    \includegraphics[width=0.48\textwidth]{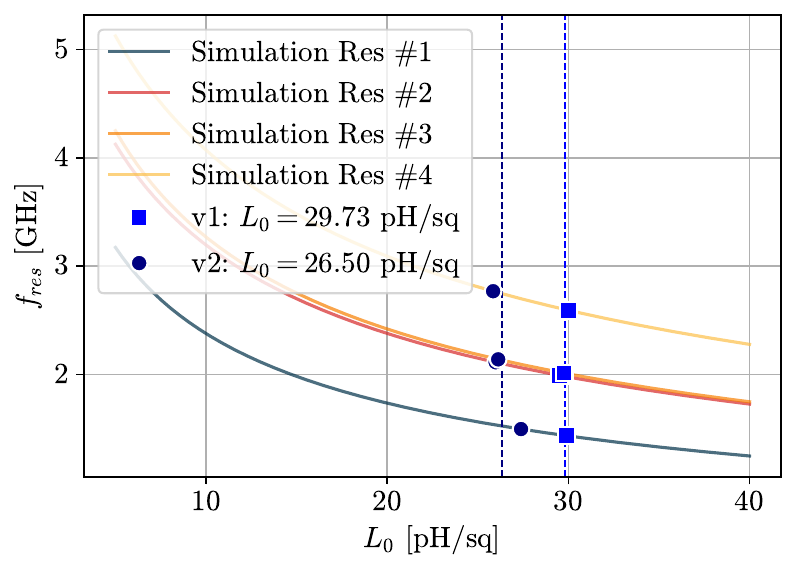}
    \caption{Example of calibration curves for the simulated resonant frequencies as a function of kinetic inductance, compared with the measured values for the different resonators. The vertical dashed black line indicates the average kinetic inductance values for different device productions (v1 and v2). The method used to extrapolate $L_k$ is the same as the one introduced in~\cite{Giachero2023characterization}.}
    \label{fig:fres_vs_Lk}
\end{figure}

\begin{table}[t] 
\centering
\caption{List of kinetic inductances and relative permittivities obtained by comparing resonator simulations with resonance frequency measurements.}
\begin{tabular}{lll}
\hline
\hline
Version \hspace{2em} & $L_k$ [pH/sq] \hspace{2em} & $\epsilon_r$ \hspace{2em} \\
\hline
v1  & 30 & 9.59 \\ 
v2  & 28 & 8.07    \\
v3  & 36 & 8.14    \\
\hline\hline
\end{tabular} 
\label{tab:perm}
\end{table}

Once $L_0$ is known is it possible to extract $\epsilon_r$ using the microwave resonator test structures. As the total area of the parallel plate capacitors in the lumped element resonator test structures is known -- and the inductance from the NbTiN segment is known from the dc measurement -- a measurement of the resonance frequencies yields the specific capacitance for the deposited a-Si on a given wafer. We note that with only this measurement we cannot break the degeneracy in the combined quantity $\epsilon_r / d$ -- where $d$ is the a-Si thickness -- since this technique only measures the parallel plate capacitor total capacitance. I.e. we are only able to measure the specific capacitance $c$ for a given wafer with this technique. 

Fig.~\ref{fig:resonance} shows a characteristic measurement of one test resonator at 100~mK. Comparing the measured resonance frequency to curves generated via simulation of the test structures while varying $\epsilon_r$ we obtain a measurement of the specific capacitance (Fig.~\ref{fig:fres_vs_eps}). In this case we assume the dielectric thickness is exactly 100~nm and obtain a-Si dielectric strengths ranging from 8.1 to 9.6 depending on the fabrication run. Using the frequency-tunable resonator measurements, and assuming the dielectric measured via the fixed resonators, we obtain our secondary measurement of $L_0$ by comparing to simulations of the resonance frequency as a function of $L_0$ (Fig.~\ref{fig:fres_vs_Lk}). In fact, we may trace out a small portion of the simulation curve by increasing the resonator dc bias (not shown) to provide additional validation of our procedures. Table~\ref{tab:perm} shows the summary of these test structure measurements across the three device versions.

\section{Complete Noise Measurement Schematic}
The detailed schematic showing the entire experimental setup in the dilution refrigerator for measuring system noise with a KIT FSA using an SNTJ calibrated noise source is shown in Fig.~\ref{fig:full_noise_schematic}. To eliminate the interference of ground loops in accurately determining and setting the SNTJ bias we float the SNTJ box and bias tee case (rf ground) from the fridge ground using an inner/outer dc block and cables with a broken outer shield (gray line in diagram) -- necessary because the SNTJ base electrode is shorted to rf ground and thus the entire cryostat body. Additionally, we use SMA tees to enable a four-wire, in-situ, resistance measurement of the SNTJ while cold. Both these techniques improve the accuracy to which the SNTJ bias, and its corresponding output white noise level, may be known.

\section{Transmittivity Determination}\label{app:etas}
Following the analysis in Sec.~\ref{sec:noise} it is important to obtain quality estimates of the multi-component transmittivities between the output of the SNTJ itself, and the input to the KIT amplification medium -- i.e. between the reference planes on-chip and inside the SNTJ and KIT packages where signal generation and amplification occurs / begins. Recall in Fig.~\ref{fig:noise_meas}(a) we make a distinction between loss incurred by components required for the STNJ operation ($\eta_0$) and for KIT operation ($\eta_1$). 

\begin{figure*}[t]
    \centering
    \includegraphics[width = 0.98\textwidth]{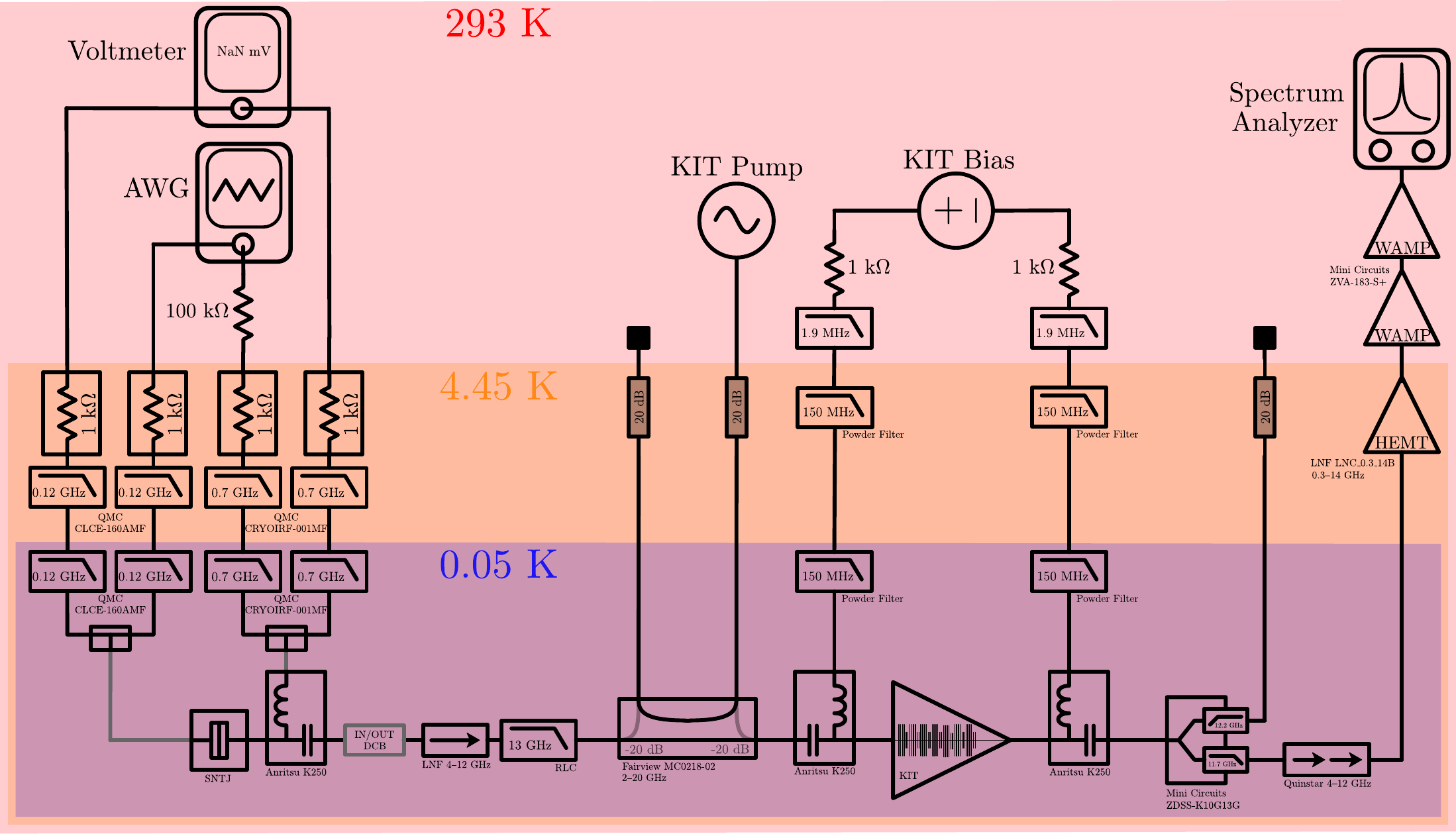}
    \caption{Complete dilution refrigerator setup schematic for the SNTJ system noise measurement. The KIT pump and VNA probe are injected at the KIT input using a -20~dB directional coupler at 50~mK. All connections both outside and inside the cryostat are made using coaxial cables; including dc bias lines. Connections indicated in gray (SNTJ bias and voltage leads) are made using cables where continuity in the outer cable shield is broken to permit differential biasing of the SNTJ.}
    \label{fig:full_noise_schematic}
\end{figure*}

\begin{figure}[b]
    \centering
    \includegraphics[width=0.48\textwidth]{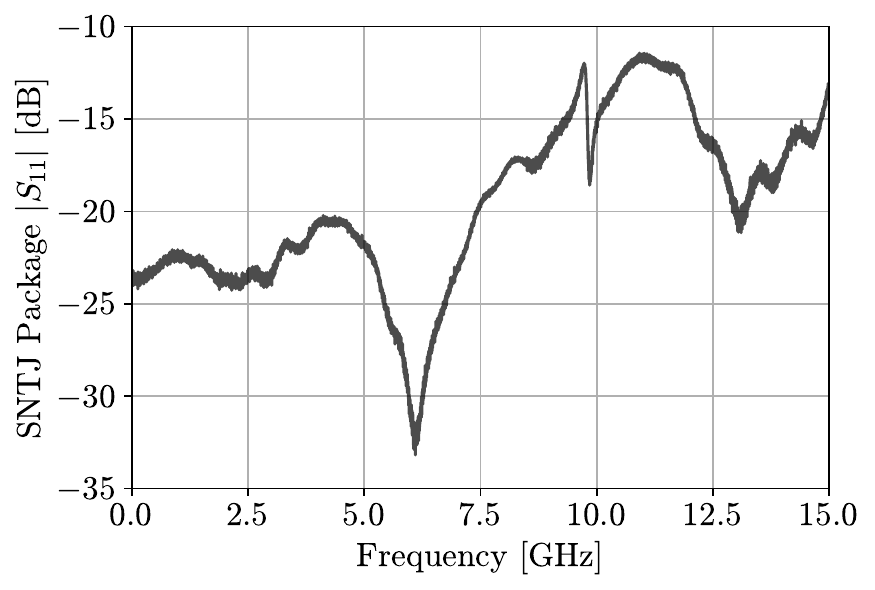}
    \caption{Return loss of the SNTJ mounted in its package and measured at room temperature. In determining $\eta_0$ we are forced to assume this does not change appreciably upon cooldown to 50~mK as we do not have the infrastructure to perform a calibrated measurement inside the dilution refrigerator while cold. Since the SNTJ impedance is consistent at room temperature and 50~mK this is a good assumption.}
    \label{fig:sntj_s11_300K}
\end{figure}

Using the four-wire technique we obtain an SNTJ (dc) impedance of $Z_0 = 47.63~\Omega$ at both room temperature and 50~mK. The reflection coefficient for this impedance is $\Gamma = -30$~dB so it is unlikely the transmittivity between the SNTJ and the KIT amplification medium is dominated by this match. Instead, we measure the return loss of the SNTJ and its package at room temperature, as well as individual component transmittivities at 50~mK. The return loss of the SNTJ-package at room temperature is shown in Fig.~\ref{fig:sntj_s11_300K} which indicates the transmittivity of the SNTJ-package composite is bounded below at $\eta_\text{SNTJ} \geq 0.94$ by the $|S_{11}| \sim -12$~dB feature between 9.5~GHz and 12~GHz. As we discuss below, the measured insertion loss of the discrete components dominates the total transmittivity.

From Fig.~\ref{fig:full_noise_schematic} we see $\eta_0$ is comprised of a dc block, a single-stage cryogenic isolator, and a lowpass filter. We measure the individual component transmittivities using paired SP6T cryogenic MEMS switches with matched cable lengths, and an SMA female-female barrel as the unity transmittivity reference, to remove all ingress/egress cabling loss. Over the KIT gain bandwidth our component responses are flat so we quote only the mean insertion loss. The dc block insertion loss is not measurable to within our measurement precision and within the uniformity of the different paths in the MEMS switches -- so we choose -0.1~dB as a conservative upper bound. The isolator and lowpass filter have an insertion loss of -0.3~dB and -0.2~dB, respectively, giving $\eta_0 = 0.93$. The transmittivity of the KIT components, a directional coupler and bias tee, are measured to be -0.25~dB and -0.2~dB -- therefore $\eta_1 = 0.95$.

\end{document}